\documentclass[a4paper,11pt]{article}

\usepackage{hyperref}%
\hypersetup{%
  colorlinks=true,%
  linkcolor=blue,%
  citecolor=blue,%
  urlcolor=black,%
  bookmarksnumbered=true,%
  bookmarkstype=toc,%
  bookmarksopen=false,%
  pdftitle={A Generalization of the One-Dimensional Boson-Fermion
    Duality Through the Path-Integral Formalsim},%
  pdfauthor={Satoshi Ohya}}%

\usepackage[margin=1in]{geometry}%

\usepackage[tt=false,sb]{libertine}%
\usepackage[T1]{fontenc}%
\usepackage{libertinust1math}%
\usepackage[cal=stix,scr=boondoxo,bb=boondox]{mathalpha}%
\usepackage{bm}%
\allowdisplaybreaks[1]%

\usepackage{epic,eepic,xcolor}%

\usepackage[font=small,hang]{caption}%
\usepackage{sidecap,subcaption}%

\usepackage[backend=biber,style=phys,biblabel=brackets,block=ragged,eprint=true,url=true]{biblatex}
\title{\Large\bfseries
  A Generalization of the One-Dimensional Boson-Fermion Duality\\
  Through the Path-Integral Formalism}%
\author{\normalsize Satoshi Ohya\\[1em]
  \small\itshape Institute of Quantum Science, Nihon University,\\
  \small\itshape Kanda-Surugadai 1-8-14, Chiyoda, Tokyo 101-8308, Japan\\[1ex]
  \small\ttfamily ohya.satoshi@nihon-u.ac.jp}%
\date{\small(Dated: \today)}%

\begin{document}
\maketitle%
\flushbottom%

\begin{abstract}
  We study boson-fermion dualities in one-dimensional many-body
  problems of identical particles interacting only through two-body
  contacts. By using the path-integral formalism as well as the
  configuration-space approach to indistinguishable particles, we find
  a generalization of the boson-fermion duality between the
  Lieb-Liniger model and the Cheon-Shigehara model. We present an
  explicit construction of $n$-boson and $n$-fermion models which are
  dual to each other and characterized by $n-1$ distinct
  (coordinate-dependent) coupling constants. These models enjoy the
  spectral equivalence, the boson-fermion mapping, and the strong-weak
  duality. We also discuss a scale-invariant generalization of the
  boson-fermion duality.
\end{abstract}

\newpage
\section{Introduction}
\label{section:1}
In his seminal paper \cite{Girardeau:1960} in 1960, Girardeau proved
the one-to-one correspondence---the duality---between one-dimensional
spinless bosons and fermions with hard-core interparticle
interactions. By using this duality, he presented a celebrated example
of the spectral equivalence between impenetrable bosons and free
fermions. Since then, the one-dimensional boson-fermion duality has
been a testing ground for studying strongly-interacting many-body
problems, especially in the field of integrable models. So far there
have been proposed several generalizations of the Girardeau's finding,
the most prominent of which was given by Cheon and Shigehara in 1998
\cite{Cheon:1998iy}: they discovered the fermionic dual of the
Lieb-Liniger model \cite{Lieb:1963rt} by using the generalized
pointlike interactions. The duality between the Lieb-Liniger model and
the Cheon-Shigehara model is a natural generalization of
\cite{Girardeau:1960} and consists of (i) the spectral equivalence
between the bosonic and fermionic systems, (ii) the one-to-one mapping
between bosonic and fermionic wavefunctions, and (iii) the one-to-one
correspondence between a strong-coupling regime in one system and a
weak-coupling regime in the other (i.e., the strong-weak duality). The
purpose of the present paper is to derive and further generalize this
duality by using the path-integral formalism. Before going into
details, however, let us first briefly recall the basics of the
boson-fermion duality by simplifying the argument of
\cite{Cheon:1998iy}.

Let us consider $n$ identical particles which have no internal
structures (i.e., spinless), move on the whole line $\mathbb{R}$, and
interact through two-body contact interactions but otherwise freely
propagate in the bulk. The Lieb-Liniger model and the Cheon-Shigehara
model are particular examples of such $n$-body systems and described
by the following Hamiltonians, respectively:
\begin{subequations}
  \begin{align}
    H_{\text{B}}&=-\frac{\hbar^{2}}{2m}\sum_{j=1}^{n}\frac{\partial^{2}}{\partial x_{j}^{2}}+\frac{\hbar^{2}}{m}\sum_{1\leq j<k\leq n}\delta(x_{j}-x_{k};\tfrac{1}{a}),\label{eq:1a}\\
    H_{\text{F}}&=-\frac{\hbar^{2}}{2m}\sum_{j=1}^{n}\frac{\partial^{2}}{\partial x_{j}^{2}}+\frac{\hbar^{2}}{m}\sum_{1\leq j<k\leq n}\varepsilon(x_{j}-x_{k};a),\label{eq:1b}
  \end{align}
\end{subequations}
where $m$ is the mass of the particles and $x_{j}$ is the coordinate
of the $j$th particle. Here
$\delta(x;\tfrac{1}{a})=\tfrac{1}{a}\delta(x)$ is the
$\delta$-function potential with $a$ being a real constant that has
the dimension of length and $\varepsilon(x;a)$ is the so-called
$\varepsilon$-function potential defined by a limit of a particular
linear combination of the $\delta$-functions \cite{Cheon:1997rx}. To
see the duality, however, there is no need to know the precise
definition of $\varepsilon(x;a)$ because both the $\delta$- and
$\varepsilon$-function potentials are prescribed by connection
conditions for wavefunctions. In other words, the $n$-body systems
described by the Hamiltonians \eqref{eq:1a} and \eqref{eq:1b} are
equivalently described by the bulk free Hamiltonian
$H_{0}=-\frac{\hbar^{2}}{2m}\sum_{j=1}^{n}\frac{\partial^{2}}{\partial
  x_{j}^{2}}$ with some connection conditions at the coincidence
points $x_{j}=x_{k}$. For example, the $n$-body Hamiltonian
\eqref{eq:1a} is described by the bulk free Hamiltonian $H_{0}$ with
the following connection conditions for the $\delta$-function
potential $\delta(x_{j}-x_{k};\tfrac{1}{a})$:
\begin{subequations}
  \begin{align}
    \frac{\partial\psi_{\text{B}}}{\partial x_{jk}}\biggr|_{x_{jk}=0_{+}}-\frac{\partial\psi_{\text{B}}}{\partial x_{jk}}\biggr|_{x_{jk}=0_{-}}-\frac{1}{a}\left(\psi_{\text{B}}\bigr|_{x_{jk}=0_{+}}+\psi_{\text{B}}\bigr|_{x_{jk}=0_{-}}\right)&=0,\label{eq:2a}\\
    \psi_{\text{B}}\bigr|_{x_{jk}=0_{+}}-\psi_{\text{B}}\bigr|_{x_{jk}=0_{-}}&=0,\label{eq:2b}
  \end{align}
\end{subequations}
where $x_{jk}=x_{j}-x_{k}$ and
$\frac{\partial}{\partial x_{jk}}=\frac{\partial}{\partial
  x_{j}}-\frac{\partial}{\partial x_{k}}$. On the other hand, the
$n$-body Hamiltonian \eqref{eq:1b} is described by the bulk free
Hamiltonian $H_{0}$ with the following connection conditions for the
$\varepsilon$-function potential $\varepsilon(x_{j}-x_{k};a)$
\cite{Cheon:1998iy,Cheon:1997rx}:\footnote{Yet another
  pseudo-potential realization of the connection conditions
  \eqref{eq:3a} and \eqref{eq:3b} was discussed in
  \cite{Girardeau:2004}. Note that the contact interaction described
  by \eqref{eq:3a} and \eqref{eq:3b} is also widely called the
  ``$\delta^{\prime}$-interaction'' in the literature; see, e.g.,
  \cite{Albeverio:1988}.}
\begin{subequations}
  \begin{align}
    \psi_{\text{F}}\bigr|_{x_{jk}=0_{+}}-\psi_{\text{F}}\bigr|_{x_{jk}=0_{-}}-a\left(\frac{\partial\psi_{\text{F}}}{\partial x_{jk}}\biggr|_{x_{jk}=0_{+}}+\frac{\partial\psi_{\text{F}}}{\partial x_{jk}}\biggr|_{x_{jk}=0_{-}}\right)&=0,\label{eq:3a}\\
    \frac{\partial\psi_{\text{F}}}{\partial x_{jk}}\biggr|_{x_{jk}=0_{+}}-\frac{\partial\psi_{\text{F}}}{\partial x_{jk}}\biggr|_{x_{jk}=0_{-}}&=0.\label{eq:3b}
  \end{align}
\end{subequations}
Notice that $\psi_{\text{B}}$ and its derivatives become continuous in
the limit $1/a\to0$, which indicates that $1/a$ describes the
deviation from the smooth continuous theory (i.e., the free theory)
and plays the role of a coupling constant in the Lieb-Liniger model
\eqref{eq:1a}. In contrast, $\psi_{\text{F}}$ and its derivatives
become continuous in the limit $a\to0$, which indicates that $a$
describes the deviation from the free theory and plays the role of a
coupling constant in the Cheon-Shigehara model \eqref{eq:1b}. This
inverse relation of the coupling constants is already incorporated
into the notations $\delta(x;\tfrac{1}{a})$ and $\varepsilon(x;a)$ and
the heart of the strong-weak duality.

Now, all the above connection conditions are valid for generic
wavefunctions without any symmetry. However, further simplifications
occur if the wavefunctions are totally symmetric (antisymmetric) under
the exchange of coordinates, which must hold for identical spinless
bosons (fermions) due to the indistinguishability of identical
particles in quantum mechanics. For example, if $\psi_{\text{B}}$ is
the totally symmetric function and satisfies the identity
$\psi_{\text{B}}(\cdots,x_{j},\cdots,x_{k},\cdots)=\psi_{\text{B}}(\cdots,x_{k},\cdots,x_{j},\cdots)$,
there automatically hold the additional conditions
$\psi_{\text{B}}|_{x_{jk}=0_{+}}=\psi_{\text{B}}|_{x_{jk}=0_{-}}$ and
$\frac{\partial\psi_{\text{B}}}{\partial
  x_{jk}}\bigl|_{x_{jk}=0_{+}}=-\frac{\partial\psi_{\text{B}}}{\partial
  x_{jk}}\bigl|_{x_{jk}=0_{-}}$, which reduce \eqref{eq:2a} to
$\frac{\partial\psi_{\text{B}}}{\partial
  x_{jk}}\bigr|_{x_{jk}=0_{+}}-\frac{1}{a}\psi_{\text{B}}|_{x_{jk}=0_{+}}=0$. Similarly,
if $\psi_{\text{F}}$ is the totally antisymmetric function and
satisfies the identity
$\psi_{\text{F}}(\cdots,x_{j},\cdots,x_{k},\cdots)=-\psi_{\text{F}}(\cdots,x_{k},\cdots,x_{j},\cdots)$,
there automatically hold the additional conditions
$\psi_{\text{F}}|_{x_{jk}=0_{+}}=-\psi_{\text{F}}|_{x_{jk}=0_{-}}$ and
$\frac{\partial\psi_{\text{F}}}{\partial
  x_{jk}}\bigl|_{x_{jk}=0_{+}}=\frac{\partial\psi_{\text{F}}}{\partial
  x_{jk}}\bigl|_{x_{jk}=0_{-}}$, which reduce \eqref{eq:3a} to
$\psi_{\text{F}}|_{x_{jk}=0_{+}}-a\frac{\partial\psi_{\text{F}}}{\partial
  x_{jk}}\bigr|_{x_{jk}=0_{+}}=0$. Putting all these things together,
one immediately sees that the systems of $n$ identical bosons and
fermions described by \eqref{eq:1a} and \eqref{eq:1b} are both
described by the bulk free Hamiltonian $H_{0}$ together with the
following Robin boundary conditions at the coincidence points:
\begin{align}
  \frac{\partial\psi_{\text{B/F}}}{\partial x_{jk}}\biggr|_{x_{jk}=0_{+}}-\frac{1}{a}\psi_{\text{B/F}}\bigr|_{x_{jk}=0_{+}}=0.\label{eq:4}
\end{align}
Since both systems are described by the same bulk Hamiltonian and the
same boundary conditions, they automatically become isospectral. This
is the boson-fermion duality in one dimension, which consists of the
spectral equivalence between $H_{\text{B}}$ and $H_{\text{F}}$, the
equivalence between the strong coupling regime of $H_{\text{B/F}}$ and
the weak coupling regime of $H_{\text{F/B}}$, and, as we will see in
section \ref{section:2.3}, the one-to-one mapping between
$\psi_{\text{B}}$ and $\psi_{\text{F}}$.  Note that the extreme case
$a\to0$ corresponds to the simplest duality between the impenetrable
bosons and the free fermions.

Now it is obvious from the above discussion that the one-dimensional
boson-fermion duality just follows from simple connection condition
arguments. However, it took more than thirty years to arrive at the
above findings since the discovery of the simplest duality by
Girardeau. One reason for this would be the lack of a systematic
derivation of the dual contact interactions for identical bosons and
fermions. The purpose of the present paper is to fill this gap and to
present a systematic machinery for deriving (and generalizing) the
boson-fermion duality. As we will see in the rest of the paper, this
purpose is achieved by using the path-integral formalism as well as
the configuration-space approach to identical particles
\cite{Souriau:1967,Souriau:1969,Laidlaw:1970ei,Leinaas:1977fm}.

The paper is organized as follows. In section \ref{section:2} we first
present the basics of the configuration-space approach to identical
particles following the argument of Leinaas and Myrheim
\cite{Leinaas:1977fm}. In this approach, the indistinguishability of
identical particles is incorporated into the configuration space
rather than the permutation symmetry of multiparticle
wavefunctions. We first see that the configuration space of $n$
identical particles on $\mathbb{R}$ is given by the orbit space
$\mathcal{M}_{n}=(\mathbb{R}^{n}-\Delta_{n})/S_{n}$, where
$\Delta_{n}$ is the set of coincidence points of two or more particles
and $S_{n}$ the symmetric group. We see that this space has a number
of nontrivial boundaries, where $k$-body contact interactions take
place at codimension-$(k-1)$ boundaries. We then see that,
irrespective of the particle statistics, two-body contact interactions
are generally described by the Robin boundary conditions at the
codimension-$1$ boundaries of $\mathcal{M}_{n}$. We also discuss that
the boson-fermion mapping holds irrespective of the boundary
conditions. In section \ref{section:3} we study the Feynman kernel for
identical particles from the viewpoint of path integral. We will show
that, by using the Feynman kernel on $\mathcal{M}_{n}$, the
boson-fermion duality between \eqref{eq:1a} and \eqref{eq:1b} is
generalized to the duality between the models described by the
Hamiltonians \eqref{eq:49a} and \eqref{eq:49b}. In section
\ref{section:4} we summarize our results with possible future
directions and discuss a scale-invariant generalization of the
boson-fermion duality. Appendices \ref{appendix:A} and
\ref{appendix:B} present proofs of some mathematical formulae.

\section{Identical particles in one dimension}
\label{section:2}
One of the principles of quantum mechanics is the indistinguishability
of identical particles. There are two main approaches to implement
this principle into a theory. The first is to consider the permutation
symmetry of multiparticle wavefunctions. The second is to restrict the
multiparticle configuration space by identifying all the permuted
points. As we will see below, the one-dimensional boson-fermion
duality is best described by the second approach. In order to fix the
notations, however, let us first start with the first approach.

Let us consider $n$ identical particles moving on the whole line
$\mathbb{R}$. Let us label each particle with a number
$j\in\{1,\cdots,n\}$ and let $x_{j}\in\mathbb{R}$ be a spatial
coordinate of the $j$th particle. Let $\sigma\in S_{n}$ be a
permutation of $n$ indices that acts on a multiparticle wavefunction
$\psi(\bm{x})=\psi(x_{1},\cdots,x_{n})$ as follows:
\begin{align}
  \sigma:\psi(\bm{x})\mapsto\psi(\sigma\bm{x}),\label{eq:5}
\end{align}
where $\sigma\bm{x}$ is defined as
$\sigma\bm{x}:=(x_{\sigma(1)},\cdots,x_{\sigma(n)})$. Note that this
definition satisfies the identity
$\sigma(\sigma^{\prime}\bm{x})=(\sigma\sigma^{\prime})\bm{x}\,(=(x_{\sigma(\sigma^{\prime}(1))},\cdots,x_{\sigma(\sigma^{\prime}(n))}))$
for any $\sigma,\sigma^{\prime}\in S_{n}$. The indistinguishability of
identical particles then implies that the multiparticle wavefunctions
$\psi(\bm{x})$ and $\psi(\sigma\bm{x})$ are physically equivalent;
that is, the probability densities for these configurations must be
the same. Thus we have
\begin{align}
  |\psi(\sigma\bm{x})|^{2}=|\psi(\bm{x})|^{2}.\label{eq:6}
\end{align}
In other words, $\psi(\sigma\bm{x})$ and $\psi(\bm{x})$ must be
identical up to a phase factor. Hence,
\begin{align}
  \psi(\sigma\bm{x})=\chi(\sigma)\psi(\bm{x}),\label{eq:7}
\end{align}
where $\chi(\sigma)\in U(1)$ is a phase that may depend on
$\sigma$.\footnote{We here assume that $\chi(\sigma)$ is independent
  of the coordinates. This assumption excludes, for example, the anyon
  exchange relations in \cite{Kundu:1998mp,Girardeau:2006}. For
  simplicity, we will not touch upon coordinate-dependent
  particle-exchange phases in the present paper. Note that a
  path-integral approach to one-dimensional anyons was discussed in
  \cite{Zhu:2007}.}  In addition, the identity
$\psi(\sigma(\sigma^{\prime}\bm{x}))=\psi((\sigma\sigma^{\prime})\bm{x})$
implies that $\chi$ must preserve the group multiplication law
$\chi(\sigma)\chi(\sigma^{\prime})=\chi(\sigma\sigma^{\prime})$ for
any $\sigma,\sigma^{\prime}\in S_{n}$; that is, the map
$\chi:S_{n}\mapsto U(1)$ must be a one-dimensional unitary
representation of $S_{n}$. As is well-known, there are just two such
representations of $S_{n}$, one is the totally symmetric
representation (i.e., the trivial representation) $\chi^{\text{[B]}}$
and the other the totally antisymmetric representation (i.e., the sign
representation) $\chi^{\text{[F]}}$, both of which are simply given by
\begin{subequations}
  \begin{align}
    \chi^{\text{[B]}}(\sigma)&=1,\label{eq:8a}\\
    \chi^{\text{[F]}}(\sigma)&=\operatorname{sgn}(\sigma).\label{eq:8b}
  \end{align}
\end{subequations}
Here $\operatorname{sgn}(\sigma)$ stands for the signature of $\sigma$
and is given by $\operatorname{sgn}(\sigma)=1$ for even permutations
and $\operatorname{sgn}(\sigma)=-1$ for odd permutations. Of course,
the totally symmetric representation $\chi^{\text{[B]}}$ corresponds
to the Bose-Einstein statistics and the totally antisymmetric
representation $\chi^{\text{[F]}}$ the Fermi-Dirac statistics. In this
way, the particle statistics is determined by the one-dimensional
unitary representation of the symmetric group $S_{n}$.

There is another, more geometrical approach to identical particles,
which was introduced independently by Souriau
\cite{Souriau:1967,Souriau:1969} and by Laidlaw and DeWitt
\cite{Laidlaw:1970ei} and then thoroughly investigated by Leinaas and
Myrheim \cite{Leinaas:1977fm} (see also the introduction of
\cite{Harrison:2010} for a nice pedagogical review). In this approach,
the indistinguishability of identical particles is built into the
configuration space by identifying all the permuted configurations of
identical particles. More precisely, given a one-particle
configuration space $X$, the configuration space of $n$ identical
particles is generally given by first taking the Cartesian product of
$n$ copies of $X$, and then subtracting the coincidence points of two
or more particles\footnote{This subtraction procedure is the heart of
  the configuration-space approach and is based on the assumption that
  two or more particles cannot occupy the same point simultaneously
  \cite{Laidlaw:1970ei}. This assumption may be arguable, but it
  successfully leads to the braid-group statistics in two dimensions
  and the Bose-Fermi alternative in three and higher dimensions.}, and
then identifying points under the action of every permutation. The
resulting space is the orbit space
$\mathcal{M}_{n}=(X^{n}-\Delta_{n})/S_{n}$, where $X^{n}$ is the
Cartesian product of $X$ and $\Delta_{n}$ the set of coincidence
points. In this setting, a particle exchange is no longer described by
the permutation symmetry of multiparticle wavefunctions because all
the permuted configurations are identified in
$\mathcal{M}_{n}$. Instead, it is described by \textit{dynamics}:
identical particles are said to be exchanged if they start from an
initial point in $\mathcal{M}_{n}$ and then return to the same point
in $\mathcal{M}_{n}$ in the course of the time-evolution. Clearly,
such an exchange process corresponds to a closed loop in the
configuration space. In addition, if $\mathcal{M}_{n}$ is a
multiply-connected space, multiparticle wavefunctions may not return
to itself but rather acquire a phase after completing the loop. It is
such a phase that determines the particle statistics. Moreover, it can
be shown that such a phase must be a member of a one-dimensional
unitary representation of the fundamental group
$\pi_{1}(\mathcal{M}_{n})$. A typical example is the case
$X=\mathbb{R}^{d}$ with $d\geq3$, in which the fundamental group is
$S_{n}$. Therefore, in three and higher dimensions, particle-exchange
phases must be $+1$ or $-1$, thus reproducing the Bose-Fermi
alternative. Another typical example is the case $X=\mathbb{R}^{2}$,
in which the fundamental group becomes the braid group $B_{n}$. Since
there is a one-parameter family of one-dimensional unitary
representations of $B_{n}$, the particle-exchange phase must be of the
form $\mathrm{e}^{i\theta}$, thus predicting anyons that interpolate
bosons and fermions. We note that, though the particle exchange is a
dynamical process in the configuration-space approach, the particle
statistics itself is still kinematical in the sense that it is
determined by the representation theory of $\pi_{1}(\mathcal{M}_{n})$.

In one dimension, however, the situation is rather different: if
$X=\mathbb{R}$, the configuration space becomes a simply-connected
convex set with boundary (see section \ref{section:2.1}) such that any
closed loops become homotopically equivalent. One might therefore
think that multiparticle wavefunctions would not acquire any
nontrivial phase under the particle exchange and there would arise
only the trivial statistics (i.e., the Bose-Einstein statistics). This
is, however, not the case because---as we will see in the
path-integral formalism---identical particles still acquire a
nontrivial phase every time multiparticle trajectories hit the
boundaries, just as in the case of a single particle on the half-line
\cite{Clark:1980xt,Farhi:1989jz} or in a box
\cite{Janke:1979fv,Inomata:1980th,Goodman:1981} (see also
\cite{Kleinert:2009} for a textbook exposition). In addition, such a
phase turns out to be a member of a one-dimensional unitary
representation of the symmetric group $S_{n}$, thus reproducing the
Bose-Fermi alternative again in one dimension. We will revisit these
things in section \ref{section:3}.

The rest of this section is devoted to a detailed analysis of the
configuration space for $n$ identical particles on $\mathbb{R}$. We
see that a coincidence point of $k+1$ particles corresponds to a
codimension-$k$ boundary of the configuration space. We then discuss
that two-body contact interactions are generally described by the
Robin boundary conditions at the codimension-1 boundaries. Since there
are $n-1$ such boundaries, it is possible to introduce $n-1$ distinct
coupling constants in the $n$-body problem of identical particles in
one dimension. Furthermore, these coupling constants turn out to be
able to depend on the coordinates. These are in stark contrast to the
boundary conditions \eqref{eq:4} and lead us to generalize the
boson-fermion duality between the Lieb-Liniger and Cheon-Shigehara
models. Finally, we discuss the boson-fermion mapping in our setup.

\subsection{Configuration space of identical particles}
\label{section:2.1}
Let us begin with the definition of the $n$-body configuration space
in one dimension. As noted before, the configuration space of $n$
identical particles is given by first taking the Cartesian product of
the one-particle configuration space, and then subtracting the
coincidence points, and then identifying all the permuted points. The
resulting space in one dimension is the following orbit space:
\begin{align}
  \mathcal{M}_{n}=\mathring{\mathbb{R}}^{n}/S_{n},\label{eq:9}
\end{align}
where $\mathring{\mathbb{R}}^{n}$ is the configuration space for $n$
distinguishable particles on $\mathbb{R}$ and given by
\begin{align}
  \mathring{\mathbb{R}}^{n}=\mathbb{R}^{n}-\Delta_{n}.\label{eq:10}
\end{align}
Here $\Delta_{n}$ is a set of coincidence points $x_{j}=x_{k}$ where
two or more particles occupy the same point. In one dimension, such a
set can be defined as the following vanishing locus of the Vandermonde
polynomial:
\begin{align}
  \Delta_{n}=\left\{(x_{1},\cdots,x_{n})\in\mathbb{R}^{n}:\prod_{1\leq j<k\leq n}(x_{j}-x_{k})=0\right\}.\label{eq:11}
\end{align}
Note that $\mathring{\mathbb{R}}^{n}$ consists of $n!$ disconnected
regions described by the ordering
$x_{\sigma(1)}>\cdots>x_{\sigma(n)}$, where $\sigma$ runs through all
possible permutations of $n$ indices. Since all these regions are
physically identified, it is sufficient to consider only a single
region, say $x_{1}>\cdots>x_{n}$. Hence the orbit space \eqref{eq:9}
can be identified with the following bounded region in
$\mathbb{R}^{n}$:
\begin{align}
  \mathcal{M}_{n}=\{(x_{1},\cdots,x_{n})\in\mathbb{R}^{n}:x_{1}>\cdots>x_{n}\}.\label{eq:12}
\end{align}
This is the configuration space of $n$ identical particles in one
dimension. Notice that this space is a convex set and hence simply
connected\footnote{It is easy to see that, if
  $\bm{x},\bm{y}\in \mathcal{M}_{n}$, then
  $(1-s)\bm{x}+s\bm{y}\in \mathcal{M}_{n}$ for any $s\in[0,1]$. Thus
  $\mathcal{M}_{n}$ is a convex set. Note that any convex set is
  simply connected.}.

\begin{figure}[t]
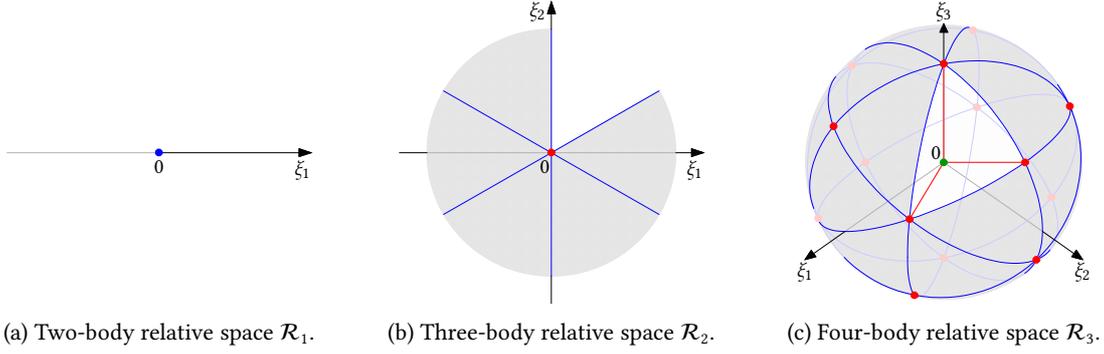

  \centering%
  \begin{subfigure}[t]{0.3\textwidth}
    \centering%
\xdefinecolor{rgb_000000}{rgb}{0,0,0}%
\xdefinecolor{rgb_0000ff}{rgb}{0,0,1}%
\xdefinecolor{rgb_b2b2b2}{rgb}{0.698039,0.698039,0.698039}%
\setlength{\unitlength}{1cm}%
\begin{picture}(4,4)(0,0)%
\color{rgb_b2b2b2}%
\path(0,2)(2,2)
\color{rgb_000000}%
\path(2,2)(4,2)
\path(3.84184,1.96046)(3.88138,1.96046)
\path(3.84184,1.97364)(3.92092,1.97364)
\path(3.84184,1.98682)(3.96046,1.98682)
\path(3.84184,2)(4,2)
\path(3.84184,2.01318)(3.96046,2.01318)
\path(3.84184,2.02636)(3.92092,2.02636)
\path(3.84184,2.03954)(3.88138,2.03954)
\path(3.98682,1.99561)(3.98682,2.00439)
\path(3.97364,1.99121)(3.97364,2.00879)
\path(3.96046,1.98682)(3.96046,2.01318)
\path(3.94728,1.98243)(3.94728,2.01757)
\path(3.9341,1.97803)(3.9341,2.02197)
\path(3.92092,1.97364)(3.92092,2.02636)
\path(3.90774,1.96925)(3.90774,2.03075)
\path(3.89456,1.96485)(3.89456,2.03515)
\path(3.88138,1.96046)(3.88138,2.03954)
\path(3.8682,1.95607)(3.8682,2.04393)
\path(3.85502,1.95167)(3.85502,2.04833)
\path(3.84184,1.94728)(3.84184,2.05272)
\path(3.84184,2)(3.84184,1.94728)(4,2)(3.84184,2.05272)(3.84184,2)
\put(2,2){\color{rgb_0000ff}$\allinethickness{0.052719cm}\circle{0.052719}$}%
\put(4,1.89456){\makebox(0,0)[tr]{\hbox{\color{rgb_000000}\scriptsize $\xi_{1}$}}}
\put(2,1.89456){\makebox(0,0)[t]{\hbox{\color{rgb_000000}\scriptsize $0$}}}
\end{picture}%
%
    \caption{Two-body relative space $\mathcal{R}_{1}$.}
    \label{figure:1a}
  \end{subfigure}\quad
  \begin{subfigure}[t]{0.3\textwidth}
    \centering%
    \input{figure1b.eepic}%
    \caption{Three-body relative space $\mathcal{R}_{2}$.}
    \label{figure:1b}
  \end{subfigure}\quad
  \begin{subfigure}[t]{0.3\textwidth}
    \centering%
    \input{figure1c.eepic}%
    \caption{Four-body relative space $\mathcal{R}_{3}$.}
    \label{figure:1c}
  \end{subfigure}
  \caption{Configuration spaces for the relative motion of two, three,
    and four identical particles in one dimension. (a)
    $\mathcal{R}_{1}=\{\xi_{1}\in\mathbb{R}:0<\xi_{1}\}$ is just the
    half-line. The blue dot represents the codimension-1 boundary at
    which a two-body contact interaction takes place. (b)
    $\mathcal{R}_{2}=\{(\xi_{1},\xi_{2})\in\mathbb{R}^{2}:0<\xi_{1}<\sqrt{3}\xi_{2}\}$
    is the infinite sector with the angle $\pi/3$. The gray shaded
    region represents the impenetrable domain for the identical
    particles. The blue lines and the red dot represent the
    codimension-1 and -2 boundaries at which two- and three-body
    contact interactions take place. (c)
    $\mathcal{R}_{3}=\{(\xi_{1},\xi_{2},\xi_{3})\in\mathbb{R}^{3}:0<\xi_{1}<\sqrt{3}\xi_{2}<\sqrt{6}\xi_{3}\}$
    is the infinite triangular pyramid. (To visualize
    $\mathcal{R}_{3}$, consider, e.g., the tetrakis hexahedron.) The
    gray shaded region represents the impenetrable domain. The blank
    white surfaces (including the blue curves), the red lines, and the
    green dot represent the codimension-1, -2, and -3 boundaries at
    which two-, three-, and four-body contact interactions take
    place.}
  \label{figure:1}
\end{figure}

Though it is not necessary for deriving the boson-fermion duality, it
may be instructive to point out here that $\mathcal{M}_{n}$ can be
factorized into the direct product of three distinct pieces---the
space of the center-of-mass motion, the space of the hyperradial
motion, and the space of the hyperangular motion. To see this, it is
convenient to introduce the following normalized Jacobi coordinates:
\begin{subequations}
  \begin{align}
    \xi_{j}&=\frac{x_{1}+\cdots+x_{j}-jx_{j+1}}{\sqrt{j(j+1)}},\quad j\in\{1,\cdots,n-1\},\label{eq:13a}\\
    \xi_{n}&=\frac{x_{1}+\cdots+x_{n}}{\sqrt{n}}.\label{eq:13b}
  \end{align}
\end{subequations}
We note that these are normalized in the sense that the coordinate
transformation $(x_{1},\cdots,x_{n})\mapsto(\xi_{1},\cdots,\xi_{n})$
is an $SO(n)$ transformation and hence preserves the dot product. Note
also that $\xi_{n}$ corresponds to the center-of-mass coordinate and
is invariant under $S_{n}$. It then follows from the definitions
\eqref{eq:13a} and \eqref{eq:13b} that there hold the identities
$\tfrac{x_{1}-x_{2}}{\sqrt{2}}=\xi_{1}$ and
$\tfrac{x_{j}-x_{j+1}}{\sqrt{2}}=-\sqrt{\frac{j-1}{2j}}\xi_{j-1}+\sqrt{\frac{j+1}{2j}}\xi_{j}$
for $j=\{2,\cdots,n-1\}$, from which one can show that the condition
$x_{1}>x_{2}>\cdots>x_{n}$ is translated into the condition
$0<\xi_{1}<\cdots<\sqrt{\tfrac{n(n-1)}{2}}\xi_{n-1}$. Note that there
is no constraint on $\xi_{n}$, meaning that the one-dimensional space
$\mathbb{R}(\ni\xi_{n})$ is factored out from $\mathcal{M}_{n}$. In
order to see further factorizations, let us next introduce the
hyperradius $r$ as follows:
\begin{align}
  r
  &=\sqrt{\xi_{1}^{2}+\cdots+\xi_{n-1}^{2}}\nonumber\\
  &=\sqrt{\bm{\xi}\cdot\bm{\xi}-\xi_{n}^{2}}\nonumber\\
  &=\sqrt{\bm{x}\cdot\bm{x}-\frac{1}{n}(x_{1}+\cdots+x_{n})^{2}}\nonumber\\
  &=\sqrt{\frac{1}{n}\sum_{1\leq j<k\leq n}(x_{j}-x_{k})^{2}},\label{eq:14}
\end{align}
where $\bm{\xi}=(\xi_{1},\cdots,\xi_{n})$,
$\bm{x}=(x_{1},\cdots,x_{n})$, and the dot stands for the dot
product. Now we write $\xi_{j}=r\Hat{\xi}_{j}$ for
$j\in\{1,\cdots,n-1\}$. Then it follows from the above discussion that
$\Hat{\xi}_{j}$ should satisfy
$\Hat{\xi}_{1}^{2}+\cdots+\Hat{\xi}_{n-1}^{2}=1$ and
$0<\Hat{\xi}_{1}<\cdots<\sqrt{\tfrac{n(n-1)}{2}}\Hat{\xi}_{n-1}$. Putting
all the above things together, we arrive at the following
factorization of the configuration space:
\begin{align}
  \mathcal{M}_{n}=\mathbb{R}\times\mathbb{R}_{+}\times\Omega_{n-2},\label{eq:15}
\end{align}
where $\mathbb{R}=\left\{\xi_{n}:-\infty<\xi_{n}<\infty\right\}$ is
the space of the center-of-mass motion,
$\mathbb{R}_{+}=\left\{r:r>0\right\}$ the space of the hyperradial
motion, and $\Omega_{n-2}$ the space of the hyperangular motion given
by
\begin{align}
  \Omega_{n-2}=\left\{(\Hat{\xi}_{1},\cdots,\Hat{\xi}_{n-1})\in\mathbb{R}^{n-1}:\Hat{\xi}_{1}^{2}+\cdots+\Hat{\xi}_{n-1}^{2}=1~~\&~~0<\Hat{\xi}_{1}<\cdots<\sqrt{\frac{n(n-1)}{2}}\Hat{\xi}_{n-1}\right\}.\label{eq:16}
\end{align}
We note that the last factor $\Omega_{n-2}$ in \eqref{eq:15} must be
discarded if $n=2$. Note also that the subspace
$\mathcal{R}_{n-1}=\mathbb{R}_{+}\times\Omega_{n-2}$\footnote{The
  relative space can also be written as
  $\mathcal{R}_{n-1}=\{(\xi_{1},\cdots,\xi_{n-1})\in\mathbb{R}^{n-1}:0<\xi_{1}<\cdots<\sqrt{\tfrac{n(n-1)}{2}}\xi_{n-1}\}$.}
is nothing but the relative space in \cite{Leinaas:1977fm} that
describes the relative motion of identical particles. Typical examples
of the relative space are depicted in figure \ref{figure:1}.

Now, as can be observed from figure \ref{figure:1}, there are a number
of nontrivial boundaries in the relative space
$\mathcal{R}_{n-1}$. For example, for $n=4$ (see figure
\ref{figure:1c}), there are (i) three codimension-1 boundaries, (ii)
three codimension-2 boundaries, and (iii) a single codimension-3
boundary, which, in the original Cartesian coordinates, correspond to
(i) $\{x_{1}=x_{2}>x_{3}>x_{4}\}$, $\{x_{1}>x_{2}=x_{3}>x_{4}\}$,
$\{x_{1}>x_{2}>x_{3}=x_{4}\}$, (ii) $\{x_{1}=x_{2}=x_{3}>x_{4}\}$,
$\{x_{1}=x_{2}>x_{3}=x_{4}\}$, $\{x_{1}>x_{2}=x_{3}=x_{4}\}$, and
(iii) $\{x_{1}=x_{2}=x_{3}=x_{4}\}$, respectively. In general, a
coincidence point of $k$ particles corresponds to one of the
codimension-$(k-1)$ boundaries of $\mathcal{R}_{n-1}$ (or
$\mathcal{M}_{n}$).

To summarize, we have seen that $k$-body contact interactions take
place at a codimension-$(k-1)$ boundary of the configuration space
$\mathcal{M}_{n}$. Since the purpose of the present paper is to derive
and generalize the dual two-body contact interactions for bosons and
fermions, in what follows we will concentrate on only the
codimension-1 boundaries.

\subsection{Two-body boundary conditions}
\label{section:2.2}
In order to construct a quantum theory on $\mathcal{M}_{n}$, we have
to specify boundary conditions for wavefunctions. Below we will do
this by imposing the probability conservation at the codimension-1
boundaries.

Let us first note that a two-body contact interaction between the
$j$th and $(j+1)$th particles takes place at the following
codimension-$1$ boundary:
\begin{align}
  \partial\mathcal{M}^{\text{2-body}}_{n,j}=\{(x_{1},\cdots,x_{n})\in\mathbb{R}^{n}:x_{1}>\cdots>x_{j}=x_{j+1}>\cdots>x_{n}\},\label{eq:17}
\end{align}
where $j\in\{1,\cdots,n-1\}$. As originally discussed by Leinaas and
Myrheim \cite{Leinaas:1977fm}, in order to ensure the probability
conservation, the normal component of the probability current density
must vanish at the boundary. Thus we impose the following condition:
\begin{align}
  \bm{n}_{j}\cdot\bm{j}=0\quad\text{on}\quad\partial \mathcal{M}^{\text{2-body}}_{n,j},\label{eq:18}
\end{align}
where $\bm{n}_{j}$ stands for a normal vector\footnote{Note that
  normal vectors become ill-defined at the codimension-$k(\geq2)$
  boundaries, because these boundaries are corner singularities in
  general; see figures \ref{figure:1b} and \ref{figure:1c}. In this
  work we will not touch upon boundary conditions at these
  singularities.} to the boundary
$\partial\mathcal{M}^{\text{2-body}}_{n,j}$ and $\bm{j}$ is the
$n$-body probability current density defined by
\begin{align}
  \bm{j}=\frac{\hbar}{2im}\left(\overline{\psi}\bm{\nabla}\psi-(\overline{\bm{\nabla}\psi})\psi\right).\label{eq:19}
\end{align}
Here $\psi=\psi(x_{1},\cdots,x_{n})$ is a multiparticle wavefunction
on $\mathcal{M}_{n}$,
$\bm{\nabla}=(\frac{\partial}{\partial
  x_{1}},\cdots,\frac{\partial}{\partial x_{n}})$ is the $n$-body
differential operator, and the overline stands for the complex
conjugate. Substituting \eqref{eq:19} into \eqref{eq:18} we get
\begin{align}
  \overline{\psi}(\bm{n}_{j}\cdot\bm{\nabla}\psi)-(\overline{\bm{n}_{j}\cdot\bm{\nabla}\psi})\psi=0
  \quad\text{on}\quad\partial\mathcal{M}^{\text{2-body}}_{n,j}.\label{eq:20}
\end{align}
Note that this is a quadratic equation of $\psi$. However, it can in
fact be linearized and enjoys a one-parameter family of solutions. As
is well-known, the solution to the equation \eqref{eq:20} is given by
the following Robin boundary condition:
\begin{align}
  \bm{n}_{j}\cdot\bm{\nabla}\psi-\frac{1}{a_{j}}\psi=0
  \quad\text{on}\quad\partial\mathcal{M}^{\text{2-body}}_{n,j},\label{eq:21}
\end{align}
where $a_{j}$ is a real parameter with the dimension of length. We
emphasize that $a_{j}$ may depend on the coordinates parallel to the
boundary. Naively, such coordinate dependence would break the
translation invariance. As originally noted in \cite{Leinaas:1977fm},
this is true for $n=2$. However, for $n\geq3$, $a_{j}$ can depend on
the coordinates without spoiling the translation invariance. We will
revisit this possibility and resulting scale- and
translation-invariant two-body contact interactions in section
\ref{section:4}.

Now, since the boundary $\partial\mathcal{M}^{\text{2-body}}_{n,j}$ is
the codimension-1 surface $x_{j}-x_{j+1}=0$ in $\mathbb{R}^{n}$, the
normal vector can be simply written as follows:
\begin{align}
  \bm{n}_{j}
  &=\bm{\nabla}(x_{j}-x_{j+1})\nonumber\\
  &=(0,\cdots,0,1,-1,0,\cdots,0),\label{eq:22}
\end{align}
from which we find that the Robin boundary condition \eqref{eq:21} is
cast into the following form:
\begin{align}
  \left(\frac{\partial}{\partial x_{j}}-\frac{\partial}{\partial x_{j+1}}\right)\psi-\frac{1}{a_{j}}\psi=0
  \quad\text{on}\quad\partial\mathcal{M}^{\text{2-body}}_{n,j}.\label{eq:23}
\end{align}
This is the boundary condition that describes the two-body contact
interaction for identical particles.

It should be noted that Leinaas and Myrheim interpreted the parameter
$a_{j}$ as a statistics parameter that interpolates the Bose-Einstein
and Fermi-Dirac statistics \cite{Leinaas:1977fm}, because $a_{j}$
continuously interpolates the Neumann boundary condition (i.e., the
boundary condition for free bosons) and the Dirichlet boundary
condition (i.e., the boundary condition for free fermions). They then
advocated the existence of intermediate statistics in one
dimension. In the present paper, however, our take is different: the
parameter $a_{j}$ just describes the two-body contact interaction
rather than the statistics. As we will see in section \ref{section:3},
for bosons, the Robin boundary condition \eqref{eq:23} is translated
into the $\delta$-function potential whose coupling constant is
$1/a_{j}$ and whose support is the codimension-1 singularity
$\{x_{1}>\cdots>x_{j}=x_{j+1}>\cdots>x_{n}\}$ in
$\mathring{\mathbb{R}}^{n}$. For fermions, on the other hand,
\eqref{eq:23} turns out to become the $\varepsilon$-function potential
whose coupling constant is $a_{j}$ and whose support is
$\{x_{1}>\cdots>x_{j}=x_{j+1}>\cdots>x_{n}\}$.

\subsection{Boson-fermion mapping}
\label{section:2.3}
Before closing this section, let us discuss here the boson-fermion
mapping in terms of multiparticle wavefunctions on
$\mathring{\mathbb{R}}^{n}$. To this end, let us first suppose that we
find a normalized wavefunction $\psi(\bm{x})$ on the region
$x_{1}>\cdots>x_{n}$ (i.e., the configuration space
$\mathcal{M}_{n}$). Let us then extend this wavefunction by
introducing the following two distinct wavefunctions on the region
$x_{\sigma(1)}>\cdots>x_{\sigma(n) }$:
\begin{subequations}
  \begin{align}
    \psi_{\text{B}}(\bm{x})&:=\frac{1}{\sqrt{n!}}\psi(\sigma\bm{x}),\label{eq:24a}\\
    \psi_{\text{F}}(\bm{x})&:=\frac{1}{\sqrt{n!}}\operatorname{sgn}(\sigma)\psi(\sigma\bm{x}).\label{eq:24b}
  \end{align}
\end{subequations}
As $\sigma$ runs through all possible permutations,
eqs.~\eqref{eq:24a} and \eqref{eq:24b} define normalized wavefunctions
on $\mathring{\mathbb{R}}^{n}$. By construction, it is obvious that
$\psi_{\text{B}}$ and $\psi_{\text{F}}$ are totally symmetric and
antisymmetric under the permutation of coordinates, thus providing the
wavefunctions of identical spinless bosons and fermions on
$\mathring{\mathbb{R}}^{n}$. It is also obvious by construction that
there holds the identity
$\psi_{\text{F}}(\bm{x})=\operatorname{sgn}(\sigma)\psi_{\text{B}}(\bm{x})$
on the region $x_{\sigma(1)}>\cdots>x_{\sigma(n)}$. An alternative
equivalent expression for this is the following identity on
$\mathring{\mathbb{R}}^{n}$:
\begin{align}
  \psi_{\text{F}}(\bm{x})=\left(\prod_{1\leq j<k\leq n}\operatorname{sgn}(x_{j}-x_{k})\right)\psi_{\text{B}}(\bm{x}),\quad\forall\bm{x}\in\mathring{\mathbb{R}}^{n},\label{eq:25}
\end{align}
where $\operatorname{sgn}(x)$ here stands for the sign function,
$\operatorname{sgn}(x)=x/|x|$. This is the celebrated boson-fermion
mapping in one dimension \cite{Girardeau:1960}. It is now clear that
the fundamental ingredient of the boson-fermion duality is the
wavefunction $\psi$ on the configuration space $\mathcal{M}_{n}$: if
$\psi_{\text{B}}$ and $\psi_{\text{F}}$ are constructed from the same
$\psi$, the bosonic and fermionic systems are automatically
isospectral. It should be emphasized that the identity \eqref{eq:25}
holds irrespective of boundary conditions.

\section{Dual description of the Feynman kernel on
  \texorpdfstring{$\mathcal{M}_{n}=\mathring{\mathbb{R}}^{n}/S_{n}$}{the
    configuration space}}
\label{section:3}
In the previous section, we have presented a detailed analysis of the
configuration space of $n$ identical particles in one dimension. Note,
however, that the particle statistics is still unclear at this stage:
it is not clear whether and how multiparticle wavefunctions on
$\mathcal{M}_{n}$ acquire a phase under the process of particle
exchange. As noted at the beginning of section \ref{section:2}, the
particle exchange is a dynamical process in the configuration-space
approach. Hence it is natural to expect that particle-exchange phases
may show up by studying dynamics. In general, dynamics in quantum
mechanics is described by the Feynman kernel---the integral kernel of
the time-evolution operator---which, as is well-known, can be studied
by the path-integral formalism. In fact, the configuration-space
approach and the path-integral formalism were known to be intimately
connected. The key was the covering-space approach to the path
integral on multiply-connected spaces, which was initiated by Schulman
\cite{Schulman:1968yv} and later generalized by Laidlaw and DeWitt
\cite{Laidlaw:1970ei} and by Dowker \cite{Dowker:1972np} (see also
\cite{Horvathy:1979qk,Berg:1981ix,Horvathy:1988vh}). In this approach,
one first constructs a multiply-connected space $Q$ as
$Q=\widetilde{Q}/\Gamma$, where $\widetilde{Q}$ is a
(simply-connected) universal covering space of $Q$ and $\Gamma$ is a
discrete subgroup of the isometry of $\widetilde{Q}$ that acts freely
on $\widetilde{Q}$ (without fixed points). Then, the path integral on
$Q$ is generally given by a weighted sum of the path integrals on
$\widetilde{Q}$ with weight factors given by a one-dimensional unitary
representation of $\Gamma$. All the weight factors are linked with
homotopically distinct paths and---since $\Gamma$ is isomorphic to the
fundamental group $\pi_{1}(Q)$---coincide with the particle-exchange
phases in the configuration-space approach. In this way, the
path-integral approach to the particle statistics was successful in
deriving the Bose-Fermi alternative for $d\geq3$ \cite{Laidlaw:1970ei}
and the braid-group statistics for $d=2$ \cite{Wu:1984hj}.

In one dimension, however, the situation is rather different
again. This is because the configuration space
$\mathcal{M}_{n}=\mathring{\mathbb{R}}^{n}/S_{n}$ is not a
multiply-connected space and does not fit into the form
$Q=\widetilde{Q}/\Gamma$. In fact, to the best of our knowledge, the
path-integral derivation of the particle statistics in one dimension
is still missing.

In this section we study the Feynman kernel for $n$ identical
particles in one dimension by using the path-integral formalism. We
will see that---though $\mathring{\mathbb{R}}^{n}$ is not a universal
covering space of $\mathcal{M}_{n}$ and $S_{n}$ is not the fundamental
group of $\mathcal{M}_{n}$ either---the covering-space-like approach
still works in one dimension: the Feynman kernel on
$\mathcal{M}_{n}=\mathring{\mathbb{R}}^{n}/S_{n}$ turns out to be
given by a weighted sum of the Feynman kernels on
$\mathring{\mathbb{R}}^{n}$ with weight factors given by a
one-dimensional unitary representation of $S_{n}$.\footnote{If one
  wants to use covering-space language, one should discard the
  subtraction procedure and consider the \textit{orbifold}
  $\breve{\mathcal{M}}_{n}=\mathbb{R}^{n}/S_{n}$
  \cite{Landsman:2016zoh}. In this case, one can say that the path
  integral on $\breve{\mathcal{M}}_{n}$ is given by a weighted sum of
  the path integrals on the \textit{orbifold universal cover}
  $\mathbb{R}^{n}$ and that weight factors are given by a
  one-dimensional unitary representation of the \textit{orbifold
    fundamental group}
  $\pi^{\text{orb}}_{1}(\breve{\mathcal{M}}_{n})\cong S_{n}$. (For
  orbifolds, see, e.g., the Thurston's lecture notes
  \cite{Thurston:2002}.)  Notice that, in one dimension, the
  difference between
  $\breve{\mathcal{M}}_{n}=\mathbb{R}^{n}/S_{n}=\{(x_{1},\cdots,x_{n}):x_{1}\geq\cdots\geq
  x_{n}\}$ and
  $\mathcal{M}_{n}=\mathring{\mathbb{R}}^{n}/S_{n}=\{(x_{1},\cdots,x_{n}):x_{1}>\cdots>x_{n}\}$
  is merely the boundary: the former includes the boundary but the
  latter does not. Hence it is almost a matter of preference which one
  we should use. In this paper we will use
  $\mathcal{M}_{n}=\mathring{\mathbb{R}}^{n}/S_{n}$ in order to make
  the conceptual transition from $d=1$ to $d\geq2$ smooth.} As we will
see shortly, this leads to the Bose-Fermi alternative as well as a
generalized form of the boson-fermion duality.

To begin with, let us consider the simplest situation where the
particles freely propagate in the bulk yet interact only through the
two-body contact interactions described by \eqref{eq:23}. The dynamics
of such a system is described by the following time-dependent
Schr\"{o}dinger equation on $\mathcal{M}_{n}$:
\begin{align}
  \left(i\hbar\frac{\partial}{\partial t}+\frac{\hbar^{2}}{2m}\bm{\nabla}^{2}\right)\psi(\bm{x},t)=0,\quad\forall\bm{x}\in\mathcal{M}_{n},\label{eq:26}
\end{align}
with the Robin boundary conditions \eqref{eq:23}. A standard approach
to solve this problem is to find out the Feynman kernel
$K_{\mathcal{M}_{n}}(\bm{x},\bm{y};t)=\langle\bm{x}|\exp(-\frac{i}{\hbar}Ht)|\bm{y}\rangle$
on $\mathcal{M}_{n}$, which is the coordinate representation of the
time-evolution operator and must satisfy the following properties:
\begin{itemize}
\item{\bfseries Property 1.~(Composition law)}
  \begin{align}
    \int_{\mathcal{M}_{n}}\!\!d\bm{z}\,K_{\mathcal{M}_{n}}(\bm{x},\bm{z};t_{1})K_{\mathcal{M}_{n}}(\bm{z},\bm{y};t_{2})=K_{\mathcal{M}_{n}}(\bm{x},\bm{y};t_{1}+t_{2}),\quad\forall\bm{x},\bm{y}\in\mathcal{M}_{n},\label{eq:27}
  \end{align}
\item{\bfseries Property 2.~(Initial condition)}
  \begin{align}
    K_{\mathcal{M}_{n}}(\bm{x},\bm{y};0)=\delta(\bm{x}-\bm{y}),\quad\forall\bm{x},\bm{y}\in \mathcal{M}_{n},\label{eq:28}
  \end{align}
\item{\bfseries Property 3.~(Unitarity)}
  \begin{align}
    \overline{K_{\mathcal{M}_{n}}(\bm{x},\bm{y};t)}=K_{\mathcal{M}_{n}}(\bm{y},\bm{x};-t),\quad\forall\bm{x},\bm{y}\in\mathcal{M}_{n},\label{eq:29}
  \end{align}
\item{\bfseries Property 4.~(Schr\"{o}dinger equation)}
  \begin{align}
    \left(i\hbar\frac{\partial}{\partial t}+\frac{\hbar^{2}}{2m}\bm{\nabla}_{\bm{x}}^{2}\right)K_{\mathcal{M}_{n}}(\bm{x},\bm{y};t)=0,\quad\forall\bm{x},\bm{y}\in\mathcal{M}_{n},\label{eq:30}
  \end{align}
\item{\bfseries Property 5.~(Two-body boundary conditions)}
  \begin{align}
    \left(\frac{\partial}{\partial x_{j}}-\frac{\partial}{\partial x_{j+1}}\right)K_{\mathcal{M}_{n}}(\bm{x},\bm{y};t)-\frac{1}{a_{j}}K_{\mathcal{M}_{n}}(\bm{x},\bm{y};t)=0,\quad\forall\bm{x}\in\partial \mathcal{M}^{\text{2-body}}_{n,j},\quad\forall\bm{y}\in\mathcal{M}_{n}.\label{eq:31}
  \end{align}
\end{itemize}
Notice that the first four properties are just the coordinate
representations of the composition law
$U_{t_{1}}U_{t_{2}}=U_{t_{1}+t_{2}}$, the initial condition $U_{0}=1$,
the unitarity $U_{t}^{\dagger}=U_{t}^{-1}=U_{-t}$, and the
Schr\"{o}dinger equation
$(i\hbar\frac{\partial}{\partial t}-H)U_{t}=0$ for the time-evolution
operator $U_{t}=\exp(-\frac{i}{\hbar}Ht)$, where $H$ stands for the
Hamiltonian of the system. Once we find such a Feynman kernel, the
solution to the time-dependent Schr\"{o}dinger equation \eqref{eq:26}
can be written as the following integral transform of the initial
wavefunction at $t=0$:
\begin{align}
  \psi(\bm{x},t)=\int_{\mathcal{M}_{n}}\!\!d\bm{y}\,K_{\mathcal{M}_{n}}(\bm{x},\bm{y};t)\psi(\bm{y},0),\quad\forall\bm{x}\in\mathcal{M}_{n}.\label{eq:32}
\end{align}
It should be emphasized that, if the Feynman kernel satisfies
\eqref{eq:31}, the wavefunction given by the integral transform
\eqref{eq:32} automatically satisfies the Robin boundary conditions
\eqref{eq:23}.

Now, as we will prove in appendix \ref{appendix:A}, the Feynman kernel
on $\mathcal{M}_{n}=\mathring{\mathbb{R}}^{n}/S_{n}$ can be
constructed in almost the same way as the Dowker's covering-space
method \cite{Dowker:1972np} and written as follows:
\begin{align}
  K_{\mathcal{M}_{n}}(\bm{x},\bm{y};t)=\sum_{\sigma\in S_{n}}\chi(\sigma)K_{\mathring{\mathbb{R}}^{n}}(\bm{x},\sigma\bm{y};t),\quad\forall\bm{x},\bm{y}\in\mathcal{M}_{n},\label{eq:33}
\end{align}
where $\chi:S_{n}\to U(1)$ is a one-dimensional unitary representation
of $S_{n}$. Here $K_{\mathring{\mathbb{R}}^{n}}$ is the Feynman kernel
on $\mathring{\mathbb{R}}^{n}$ and assumed to satisfy the following
conditions:
\begin{itemize}
\item{\bfseries Assumption 1.~(Composition law)}
  \begin{align}
    \int_{\mathring{\mathbb{R}}^{n}}\!\!d\bm{z}\,K_{\mathring{\mathbb{R}}^{n}}(\bm{x},\bm{z};t_{1})K_{\mathring{\mathbb{R}}^{n}}(\bm{z},\bm{y};t_{2})=K_{\mathring{\mathbb{R}}^{n}}(\bm{x},\bm{y};t_{1}+t_{2}),\quad\forall\bm{x},\bm{y}\in\mathring{\mathbb{R}}^{n},\label{eq:34}
  \end{align}
\item{\bfseries Assumption 2.~(Initial condition)}
  \begin{align}
    K_{\mathring{\mathbb{R}}^{n}}(\bm{x},\bm{y};0)=\delta(\bm{x}-\bm{y}),\quad\forall\bm{x},\bm{y}\in\mathring{\mathbb{R}}^{n},\label{eq:35}
  \end{align}
\item{\bfseries Assumption 3.~(Unitarity)}
  \begin{align}
    \overline{K_{\mathring{\mathbb{R}}^{n}}(\bm{x},\bm{y};t)}=K_{\mathring{\mathbb{R}}^{n}}(\bm{y},\bm{x};-t),\quad\forall\bm{x},\bm{y}\in\mathring{\mathbb{R}}^{n},\label{eq:36}
  \end{align}
\item{\bfseries Assumption 4.~(Schr\"{o}dinger equation)}
  \begin{align}
    \left(i\hbar\frac{\partial}{\partial t}+\frac{\hbar^{2}}{2m}\bm{\nabla}_{\bm{x}}^{2}\right)K_{\mathring{\mathbb{R}}^{n}}(\bm{x},\bm{y};t)=0,\quad\forall\bm{x},\bm{y}\in\mathring{\mathbb{R}}^{n},\label{eq:37}
  \end{align}
\item{\bfseries Assumption 5.~(Permutation invariance)}
  \begin{align}
    K_{\mathring{\mathbb{R}}^{n}}(\bm{x},\bm{y};t)=K_{\mathring{\mathbb{R}}^{n}}(\sigma\bm{x},\sigma\bm{y};t),\quad\forall\bm{x},\bm{y}\in\mathring{\mathbb{R}}^{n},\quad\forall\sigma\in S_{n},\label{eq:38}
  \end{align}
\end{itemize}
with some connection conditions at the vanishing locus $\Delta_{n}$ in
order to glue the $n!$ disconnected regions together. It is these
connection conditions that we wish to uncover below. Before doing
this, however, let us first present a brief derivation of the formula
\eqref{eq:33} by following the Dowker's argument \cite{Dowker:1972np}.

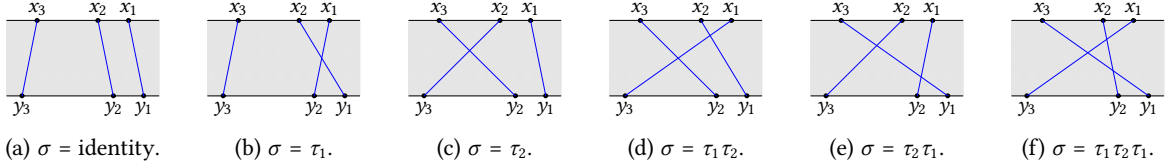
\begin{figure}[t]
  \centering%
  \begin{subfigure}[t]{0.16\textwidth}
    \centering%
\xdefinecolor{rgb_000000}{rgb}{0,0,0}%
\xdefinecolor{rgb_0000ff}{rgb}{0,0,1}%
\xdefinecolor{rgb_e5e5e5}{rgb}{0.898039,0.898039,0.898039}%
\setlength{\unitlength}{1cm}%
\begin{picture}(2,1.6)(0,0)%
\color{rgb_e5e5e5}%
\path(0,0.3)(2,0.3)
\path(0,0.313889)(2,0.313889)
\path(0,0.327778)(2,0.327778)
\path(0,0.341667)(2,0.341667)
\path(0,0.355556)(2,0.355556)
\path(0,0.369444)(2,0.369444)
\path(0,0.383333)(2,0.383333)
\path(0,0.397222)(2,0.397222)
\path(0,0.411111)(2,0.411111)
\path(0,0.425)(2,0.425)
\path(0,0.438889)(2,0.438889)
\path(0,0.452778)(2,0.452778)
\path(0,0.466667)(2,0.466667)
\path(0,0.480556)(2,0.480556)
\path(0,0.494444)(2,0.494444)
\path(0,0.508333)(2,0.508333)
\path(0,0.522222)(2,0.522222)
\path(0,0.536111)(2,0.536111)
\path(0,0.55)(2,0.55)
\path(0,0.563889)(2,0.563889)
\path(0,0.577778)(2,0.577778)
\path(0,0.591667)(2,0.591667)
\path(0,0.605556)(2,0.605556)
\path(0,0.619444)(2,0.619444)
\path(0,0.633333)(2,0.633333)
\path(0,0.647222)(2,0.647222)
\path(0,0.661111)(2,0.661111)
\path(0,0.675)(2,0.675)
\path(0,0.688889)(2,0.688889)
\path(0,0.702778)(2,0.702778)
\path(0,0.716667)(2,0.716667)
\path(0,0.730556)(2,0.730556)
\path(0,0.744444)(2,0.744444)
\path(0,0.758333)(2,0.758333)
\path(0,0.772222)(2,0.772222)
\path(0,0.786111)(2,0.786111)
\path(0,0.8)(2,0.8)
\path(0,0.813889)(2,0.813889)
\path(0,0.827778)(2,0.827778)
\path(0,0.841667)(2,0.841667)
\path(0,0.855556)(2,0.855556)
\path(0,0.869444)(2,0.869444)
\path(0,0.883333)(2,0.883333)
\path(0,0.897222)(2,0.897222)
\path(0,0.911111)(2,0.911111)
\path(0,0.925)(2,0.925)
\path(0,0.938889)(2,0.938889)
\path(0,0.952778)(2,0.952778)
\path(0,0.966667)(2,0.966667)
\path(0,0.980556)(2,0.980556)
\path(0,0.994444)(2,0.994444)
\path(0,1.00833)(2,1.00833)
\path(0,1.02222)(2,1.02222)
\path(0,1.03611)(2,1.03611)
\path(0,1.05)(2,1.05)
\path(0,1.06389)(2,1.06389)
\path(0,1.07778)(2,1.07778)
\path(0,1.09167)(2,1.09167)
\path(0,1.10556)(2,1.10556)
\path(0,1.11944)(2,1.11944)
\path(0,1.13333)(2,1.13333)
\path(0,1.14722)(2,1.14722)
\path(0,1.16111)(2,1.16111)
\path(0,1.175)(2,1.175)
\path(0,1.18889)(2,1.18889)
\path(0,1.20278)(2,1.20278)
\path(0,1.21667)(2,1.21667)
\path(0,1.23056)(2,1.23056)
\path(0,1.24444)(2,1.24444)
\path(0,1.25833)(2,1.25833)
\path(0,1.27222)(2,1.27222)
\path(0,1.28611)(2,1.28611)
\path(0,1.3)(2,1.3)
\path(2,0.3)(2,1.3)
\path(1.98601,0.3)(1.98601,1.3)
\path(1.97203,0.3)(1.97203,1.3)
\path(1.95804,0.3)(1.95804,1.3)
\path(1.94406,0.3)(1.94406,1.3)
\path(1.93007,0.3)(1.93007,1.3)
\path(1.91608,0.3)(1.91608,1.3)
\path(1.9021,0.3)(1.9021,1.3)
\path(1.88811,0.3)(1.88811,1.3)
\path(1.87413,0.3)(1.87413,1.3)
\path(1.86014,0.3)(1.86014,1.3)
\path(1.84615,0.3)(1.84615,1.3)
\path(1.83217,0.3)(1.83217,1.3)
\path(1.81818,0.3)(1.81818,1.3)
\path(1.8042,0.3)(1.8042,1.3)
\path(1.79021,0.3)(1.79021,1.3)
\path(1.77622,0.3)(1.77622,1.3)
\path(1.76224,0.3)(1.76224,1.3)
\path(1.74825,0.3)(1.74825,1.3)
\path(1.73427,0.3)(1.73427,1.3)
\path(1.72028,0.3)(1.72028,1.3)
\path(1.70629,0.3)(1.70629,1.3)
\path(1.69231,0.3)(1.69231,1.3)
\path(1.67832,0.3)(1.67832,1.3)
\path(1.66434,0.3)(1.66434,1.3)
\path(1.65035,0.3)(1.65035,1.3)
\path(1.63636,0.3)(1.63636,1.3)
\path(1.62238,0.3)(1.62238,1.3)
\path(1.60839,0.3)(1.60839,1.3)
\path(1.59441,0.3)(1.59441,1.3)
\path(1.58042,0.3)(1.58042,1.3)
\path(1.56643,0.3)(1.56643,1.3)
\path(1.55245,0.3)(1.55245,1.3)
\path(1.53846,0.3)(1.53846,1.3)
\path(1.52448,0.3)(1.52448,1.3)
\path(1.51049,0.3)(1.51049,1.3)
\path(1.4965,0.3)(1.4965,1.3)
\path(1.48252,0.3)(1.48252,1.3)
\path(1.46853,0.3)(1.46853,1.3)
\path(1.45455,0.3)(1.45455,1.3)
\path(1.44056,0.3)(1.44056,1.3)
\path(1.42657,0.3)(1.42657,1.3)
\path(1.41259,0.3)(1.41259,1.3)
\path(1.3986,0.3)(1.3986,1.3)
\path(1.38462,0.3)(1.38462,1.3)
\path(1.37063,0.3)(1.37063,1.3)
\path(1.35664,0.3)(1.35664,1.3)
\path(1.34266,0.3)(1.34266,1.3)
\path(1.32867,0.3)(1.32867,1.3)
\path(1.31469,0.3)(1.31469,1.3)
\path(1.3007,0.3)(1.3007,1.3)
\path(1.28671,0.3)(1.28671,1.3)
\path(1.27273,0.3)(1.27273,1.3)
\path(1.25874,0.3)(1.25874,1.3)
\path(1.24476,0.3)(1.24476,1.3)
\path(1.23077,0.3)(1.23077,1.3)
\path(1.21678,0.3)(1.21678,1.3)
\path(1.2028,0.3)(1.2028,1.3)
\path(1.18881,0.3)(1.18881,1.3)
\path(1.17483,0.3)(1.17483,1.3)
\path(1.16084,0.3)(1.16084,1.3)
\path(1.14685,0.3)(1.14685,1.3)
\path(1.13287,0.3)(1.13287,1.3)
\path(1.11888,0.3)(1.11888,1.3)
\path(1.1049,0.3)(1.1049,1.3)
\path(1.09091,0.3)(1.09091,1.3)
\path(1.07692,0.3)(1.07692,1.3)
\path(1.06294,0.3)(1.06294,1.3)
\path(1.04895,0.3)(1.04895,1.3)
\path(1.03497,0.3)(1.03497,1.3)
\path(1.02098,0.3)(1.02098,1.3)
\path(1.00699,0.3)(1.00699,1.3)
\path(0.993007,0.3)(0.993007,1.3)
\path(0.979021,0.3)(0.979021,1.3)
\path(0.965035,0.3)(0.965035,1.3)
\path(0.951049,0.3)(0.951049,1.3)
\path(0.937063,0.3)(0.937063,1.3)
\path(0.923077,0.3)(0.923077,1.3)
\path(0.909091,0.3)(0.909091,1.3)
\path(0.895105,0.3)(0.895105,1.3)
\path(0.881119,0.3)(0.881119,1.3)
\path(0.867133,0.3)(0.867133,1.3)
\path(0.853147,0.3)(0.853147,1.3)
\path(0.839161,0.3)(0.839161,1.3)
\path(0.825175,0.3)(0.825175,1.3)
\path(0.811189,0.3)(0.811189,1.3)
\path(0.797203,0.3)(0.797203,1.3)
\path(0.783217,0.3)(0.783217,1.3)
\path(0.769231,0.3)(0.769231,1.3)
\path(0.755245,0.3)(0.755245,1.3)
\path(0.741259,0.3)(0.741259,1.3)
\path(0.727273,0.3)(0.727273,1.3)
\path(0.713287,0.3)(0.713287,1.3)
\path(0.699301,0.3)(0.699301,1.3)
\path(0.685315,0.3)(0.685315,1.3)
\path(0.671329,0.3)(0.671329,1.3)
\path(0.657343,0.3)(0.657343,1.3)
\path(0.643357,0.3)(0.643357,1.3)
\path(0.629371,0.3)(0.629371,1.3)
\path(0.615385,0.3)(0.615385,1.3)
\path(0.601399,0.3)(0.601399,1.3)
\path(0.587413,0.3)(0.587413,1.3)
\path(0.573427,0.3)(0.573427,1.3)
\path(0.559441,0.3)(0.559441,1.3)
\path(0.545455,0.3)(0.545455,1.3)
\path(0.531469,0.3)(0.531469,1.3)
\path(0.517483,0.3)(0.517483,1.3)
\path(0.503497,0.3)(0.503497,1.3)
\path(0.48951,0.3)(0.48951,1.3)
\path(0.475524,0.3)(0.475524,1.3)
\path(0.461538,0.3)(0.461538,1.3)
\path(0.447552,0.3)(0.447552,1.3)
\path(0.433566,0.3)(0.433566,1.3)
\path(0.41958,0.3)(0.41958,1.3)
\path(0.405594,0.3)(0.405594,1.3)
\path(0.391608,0.3)(0.391608,1.3)
\path(0.377622,0.3)(0.377622,1.3)
\path(0.363636,0.3)(0.363636,1.3)
\path(0.34965,0.3)(0.34965,1.3)
\path(0.335664,0.3)(0.335664,1.3)
\path(0.321678,0.3)(0.321678,1.3)
\path(0.307692,0.3)(0.307692,1.3)
\path(0.293706,0.3)(0.293706,1.3)
\path(0.27972,0.3)(0.27972,1.3)
\path(0.265734,0.3)(0.265734,1.3)
\path(0.251748,0.3)(0.251748,1.3)
\path(0.237762,0.3)(0.237762,1.3)
\path(0.223776,0.3)(0.223776,1.3)
\path(0.20979,0.3)(0.20979,1.3)
\path(0.195804,0.3)(0.195804,1.3)
\path(0.181818,0.3)(0.181818,1.3)
\path(0.167832,0.3)(0.167832,1.3)
\path(0.153846,0.3)(0.153846,1.3)
\path(0.13986,0.3)(0.13986,1.3)
\path(0.125874,0.3)(0.125874,1.3)
\path(0.111888,0.3)(0.111888,1.3)
\path(0.0979021,0.3)(0.0979021,1.3)
\path(0.0839161,0.3)(0.0839161,1.3)
\path(0.0699301,0.3)(0.0699301,1.3)
\path(0.0559441,0.3)(0.0559441,1.3)
\path(0.041958,0.3)(0.041958,1.3)
\path(0.027972,0.3)(0.027972,1.3)
\path(0.013986,0.3)(0.013986,1.3)
\path(0,0.3)(0,1.3)
\path(0,0.3)(2,0.3)(2,1.3)(0,1.3)(0,0.3)
\color{rgb_000000}%
\path(0,0.3)(2,0.3)
\path(0,1.3)(2,1.3)
\put(1.6,1.3){\color{rgb_000000}$\allinethickness{0.035146cm}\circle{0.035146}$}%
\put(1.2,1.3){\color{rgb_000000}$\allinethickness{0.035146cm}\circle{0.035146}$}%
\put(0.4,1.3){\color{rgb_000000}$\allinethickness{0.035146cm}\circle{0.035146}$}%
\put(1.8,0.3){\color{rgb_000000}$\allinethickness{0.035146cm}\circle{0.035146}$}%
\put(1.4,0.3){\color{rgb_000000}$\allinethickness{0.035146cm}\circle{0.035146}$}%
\put(0.2,0.3){\color{rgb_000000}$\allinethickness{0.035146cm}\circle{0.035146}$}%
\color{rgb_0000ff}%
\path(1.8,0.3)(1.6,1.3)
\path(1.4,0.3)(1.2,1.3)
\path(0.2,0.3)(0.4,1.3)
\put(1.6,1.37029){\makebox(0,0)[b]{\hbox{\color{rgb_000000}\scriptsize $x_{1}$}}}
\put(1.2,1.37029){\makebox(0,0)[b]{\hbox{\color{rgb_000000}\scriptsize $x_{2}$}}}
\put(0.4,1.37029){\makebox(0,0)[b]{\hbox{\color{rgb_000000}\scriptsize $x_{3}$}}}
\put(1.8,0.229708){\makebox(0,0)[t]{\hbox{\color{rgb_000000}\scriptsize $y_{1}$}}}
\put(1.4,0.229708){\makebox(0,0)[t]{\hbox{\color{rgb_000000}\scriptsize $y_{2}$}}}
\put(0.2,0.229708){\makebox(0,0)[t]{\hbox{\color{rgb_000000}\scriptsize $y_{3}$}}}
\end{picture}%
%
    \caption{$\sigma=\text{identity}$.}
    \label{figure:2a}
  \end{subfigure}
  \begin{subfigure}[t]{0.16\textwidth}
    \centering%
\xdefinecolor{rgb_000000}{rgb}{0,0,0}%
\xdefinecolor{rgb_0000ff}{rgb}{0,0,1}%
\xdefinecolor{rgb_e5e5e5}{rgb}{0.898039,0.898039,0.898039}%
\setlength{\unitlength}{1cm}%
\begin{picture}(2,1.6)(0,0)%
\color{rgb_e5e5e5}%
\path(0,0.3)(2,0.3)
\path(0,0.313889)(2,0.313889)
\path(0,0.327778)(2,0.327778)
\path(0,0.341667)(2,0.341667)
\path(0,0.355556)(2,0.355556)
\path(0,0.369444)(2,0.369444)
\path(0,0.383333)(2,0.383333)
\path(0,0.397222)(2,0.397222)
\path(0,0.411111)(2,0.411111)
\path(0,0.425)(2,0.425)
\path(0,0.438889)(2,0.438889)
\path(0,0.452778)(2,0.452778)
\path(0,0.466667)(2,0.466667)
\path(0,0.480556)(2,0.480556)
\path(0,0.494444)(2,0.494444)
\path(0,0.508333)(2,0.508333)
\path(0,0.522222)(2,0.522222)
\path(0,0.536111)(2,0.536111)
\path(0,0.55)(2,0.55)
\path(0,0.563889)(2,0.563889)
\path(0,0.577778)(2,0.577778)
\path(0,0.591667)(2,0.591667)
\path(0,0.605556)(2,0.605556)
\path(0,0.619444)(2,0.619444)
\path(0,0.633333)(2,0.633333)
\path(0,0.647222)(2,0.647222)
\path(0,0.661111)(2,0.661111)
\path(0,0.675)(2,0.675)
\path(0,0.688889)(2,0.688889)
\path(0,0.702778)(2,0.702778)
\path(0,0.716667)(2,0.716667)
\path(0,0.730556)(2,0.730556)
\path(0,0.744444)(2,0.744444)
\path(0,0.758333)(2,0.758333)
\path(0,0.772222)(2,0.772222)
\path(0,0.786111)(2,0.786111)
\path(0,0.8)(2,0.8)
\path(0,0.813889)(2,0.813889)
\path(0,0.827778)(2,0.827778)
\path(0,0.841667)(2,0.841667)
\path(0,0.855556)(2,0.855556)
\path(0,0.869444)(2,0.869444)
\path(0,0.883333)(2,0.883333)
\path(0,0.897222)(2,0.897222)
\path(0,0.911111)(2,0.911111)
\path(0,0.925)(2,0.925)
\path(0,0.938889)(2,0.938889)
\path(0,0.952778)(2,0.952778)
\path(0,0.966667)(2,0.966667)
\path(0,0.980556)(2,0.980556)
\path(0,0.994444)(2,0.994444)
\path(0,1.00833)(2,1.00833)
\path(0,1.02222)(2,1.02222)
\path(0,1.03611)(2,1.03611)
\path(0,1.05)(2,1.05)
\path(0,1.06389)(2,1.06389)
\path(0,1.07778)(2,1.07778)
\path(0,1.09167)(2,1.09167)
\path(0,1.10556)(2,1.10556)
\path(0,1.11944)(2,1.11944)
\path(0,1.13333)(2,1.13333)
\path(0,1.14722)(2,1.14722)
\path(0,1.16111)(2,1.16111)
\path(0,1.175)(2,1.175)
\path(0,1.18889)(2,1.18889)
\path(0,1.20278)(2,1.20278)
\path(0,1.21667)(2,1.21667)
\path(0,1.23056)(2,1.23056)
\path(0,1.24444)(2,1.24444)
\path(0,1.25833)(2,1.25833)
\path(0,1.27222)(2,1.27222)
\path(0,1.28611)(2,1.28611)
\path(0,1.3)(2,1.3)
\path(2,0.3)(2,1.3)
\path(1.98601,0.3)(1.98601,1.3)
\path(1.97203,0.3)(1.97203,1.3)
\path(1.95804,0.3)(1.95804,1.3)
\path(1.94406,0.3)(1.94406,1.3)
\path(1.93007,0.3)(1.93007,1.3)
\path(1.91608,0.3)(1.91608,1.3)
\path(1.9021,0.3)(1.9021,1.3)
\path(1.88811,0.3)(1.88811,1.3)
\path(1.87413,0.3)(1.87413,1.3)
\path(1.86014,0.3)(1.86014,1.3)
\path(1.84615,0.3)(1.84615,1.3)
\path(1.83217,0.3)(1.83217,1.3)
\path(1.81818,0.3)(1.81818,1.3)
\path(1.8042,0.3)(1.8042,1.3)
\path(1.79021,0.3)(1.79021,1.3)
\path(1.77622,0.3)(1.77622,1.3)
\path(1.76224,0.3)(1.76224,1.3)
\path(1.74825,0.3)(1.74825,1.3)
\path(1.73427,0.3)(1.73427,1.3)
\path(1.72028,0.3)(1.72028,1.3)
\path(1.70629,0.3)(1.70629,1.3)
\path(1.69231,0.3)(1.69231,1.3)
\path(1.67832,0.3)(1.67832,1.3)
\path(1.66434,0.3)(1.66434,1.3)
\path(1.65035,0.3)(1.65035,1.3)
\path(1.63636,0.3)(1.63636,1.3)
\path(1.62238,0.3)(1.62238,1.3)
\path(1.60839,0.3)(1.60839,1.3)
\path(1.59441,0.3)(1.59441,1.3)
\path(1.58042,0.3)(1.58042,1.3)
\path(1.56643,0.3)(1.56643,1.3)
\path(1.55245,0.3)(1.55245,1.3)
\path(1.53846,0.3)(1.53846,1.3)
\path(1.52448,0.3)(1.52448,1.3)
\path(1.51049,0.3)(1.51049,1.3)
\path(1.4965,0.3)(1.4965,1.3)
\path(1.48252,0.3)(1.48252,1.3)
\path(1.46853,0.3)(1.46853,1.3)
\path(1.45455,0.3)(1.45455,1.3)
\path(1.44056,0.3)(1.44056,1.3)
\path(1.42657,0.3)(1.42657,1.3)
\path(1.41259,0.3)(1.41259,1.3)
\path(1.3986,0.3)(1.3986,1.3)
\path(1.38462,0.3)(1.38462,1.3)
\path(1.37063,0.3)(1.37063,1.3)
\path(1.35664,0.3)(1.35664,1.3)
\path(1.34266,0.3)(1.34266,1.3)
\path(1.32867,0.3)(1.32867,1.3)
\path(1.31469,0.3)(1.31469,1.3)
\path(1.3007,0.3)(1.3007,1.3)
\path(1.28671,0.3)(1.28671,1.3)
\path(1.27273,0.3)(1.27273,1.3)
\path(1.25874,0.3)(1.25874,1.3)
\path(1.24476,0.3)(1.24476,1.3)
\path(1.23077,0.3)(1.23077,1.3)
\path(1.21678,0.3)(1.21678,1.3)
\path(1.2028,0.3)(1.2028,1.3)
\path(1.18881,0.3)(1.18881,1.3)
\path(1.17483,0.3)(1.17483,1.3)
\path(1.16084,0.3)(1.16084,1.3)
\path(1.14685,0.3)(1.14685,1.3)
\path(1.13287,0.3)(1.13287,1.3)
\path(1.11888,0.3)(1.11888,1.3)
\path(1.1049,0.3)(1.1049,1.3)
\path(1.09091,0.3)(1.09091,1.3)
\path(1.07692,0.3)(1.07692,1.3)
\path(1.06294,0.3)(1.06294,1.3)
\path(1.04895,0.3)(1.04895,1.3)
\path(1.03497,0.3)(1.03497,1.3)
\path(1.02098,0.3)(1.02098,1.3)
\path(1.00699,0.3)(1.00699,1.3)
\path(0.993007,0.3)(0.993007,1.3)
\path(0.979021,0.3)(0.979021,1.3)
\path(0.965035,0.3)(0.965035,1.3)
\path(0.951049,0.3)(0.951049,1.3)
\path(0.937063,0.3)(0.937063,1.3)
\path(0.923077,0.3)(0.923077,1.3)
\path(0.909091,0.3)(0.909091,1.3)
\path(0.895105,0.3)(0.895105,1.3)
\path(0.881119,0.3)(0.881119,1.3)
\path(0.867133,0.3)(0.867133,1.3)
\path(0.853147,0.3)(0.853147,1.3)
\path(0.839161,0.3)(0.839161,1.3)
\path(0.825175,0.3)(0.825175,1.3)
\path(0.811189,0.3)(0.811189,1.3)
\path(0.797203,0.3)(0.797203,1.3)
\path(0.783217,0.3)(0.783217,1.3)
\path(0.769231,0.3)(0.769231,1.3)
\path(0.755245,0.3)(0.755245,1.3)
\path(0.741259,0.3)(0.741259,1.3)
\path(0.727273,0.3)(0.727273,1.3)
\path(0.713287,0.3)(0.713287,1.3)
\path(0.699301,0.3)(0.699301,1.3)
\path(0.685315,0.3)(0.685315,1.3)
\path(0.671329,0.3)(0.671329,1.3)
\path(0.657343,0.3)(0.657343,1.3)
\path(0.643357,0.3)(0.643357,1.3)
\path(0.629371,0.3)(0.629371,1.3)
\path(0.615385,0.3)(0.615385,1.3)
\path(0.601399,0.3)(0.601399,1.3)
\path(0.587413,0.3)(0.587413,1.3)
\path(0.573427,0.3)(0.573427,1.3)
\path(0.559441,0.3)(0.559441,1.3)
\path(0.545455,0.3)(0.545455,1.3)
\path(0.531469,0.3)(0.531469,1.3)
\path(0.517483,0.3)(0.517483,1.3)
\path(0.503497,0.3)(0.503497,1.3)
\path(0.48951,0.3)(0.48951,1.3)
\path(0.475524,0.3)(0.475524,1.3)
\path(0.461538,0.3)(0.461538,1.3)
\path(0.447552,0.3)(0.447552,1.3)
\path(0.433566,0.3)(0.433566,1.3)
\path(0.41958,0.3)(0.41958,1.3)
\path(0.405594,0.3)(0.405594,1.3)
\path(0.391608,0.3)(0.391608,1.3)
\path(0.377622,0.3)(0.377622,1.3)
\path(0.363636,0.3)(0.363636,1.3)
\path(0.34965,0.3)(0.34965,1.3)
\path(0.335664,0.3)(0.335664,1.3)
\path(0.321678,0.3)(0.321678,1.3)
\path(0.307692,0.3)(0.307692,1.3)
\path(0.293706,0.3)(0.293706,1.3)
\path(0.27972,0.3)(0.27972,1.3)
\path(0.265734,0.3)(0.265734,1.3)
\path(0.251748,0.3)(0.251748,1.3)
\path(0.237762,0.3)(0.237762,1.3)
\path(0.223776,0.3)(0.223776,1.3)
\path(0.20979,0.3)(0.20979,1.3)
\path(0.195804,0.3)(0.195804,1.3)
\path(0.181818,0.3)(0.181818,1.3)
\path(0.167832,0.3)(0.167832,1.3)
\path(0.153846,0.3)(0.153846,1.3)
\path(0.13986,0.3)(0.13986,1.3)
\path(0.125874,0.3)(0.125874,1.3)
\path(0.111888,0.3)(0.111888,1.3)
\path(0.0979021,0.3)(0.0979021,1.3)
\path(0.0839161,0.3)(0.0839161,1.3)
\path(0.0699301,0.3)(0.0699301,1.3)
\path(0.0559441,0.3)(0.0559441,1.3)
\path(0.041958,0.3)(0.041958,1.3)
\path(0.027972,0.3)(0.027972,1.3)
\path(0.013986,0.3)(0.013986,1.3)
\path(0,0.3)(0,1.3)
\path(0,0.3)(2,0.3)(2,1.3)(0,1.3)(0,0.3)
\color{rgb_000000}%
\path(0,0.3)(2,0.3)
\path(0,1.3)(2,1.3)
\put(1.6,1.3){\color{rgb_000000}$\allinethickness{0.035146cm}\circle{0.035146}$}%
\put(1.2,1.3){\color{rgb_000000}$\allinethickness{0.035146cm}\circle{0.035146}$}%
\put(0.4,1.3){\color{rgb_000000}$\allinethickness{0.035146cm}\circle{0.035146}$}%
\put(1.8,0.3){\color{rgb_000000}$\allinethickness{0.035146cm}\circle{0.035146}$}%
\put(1.4,0.3){\color{rgb_000000}$\allinethickness{0.035146cm}\circle{0.035146}$}%
\put(0.2,0.3){\color{rgb_000000}$\allinethickness{0.035146cm}\circle{0.035146}$}%
\color{rgb_0000ff}%
\path(1.8,0.3)(1.2,1.3)
\path(1.4,0.3)(1.6,1.3)
\path(0.2,0.3)(0.4,1.3)
\put(1.6,1.37029){\makebox(0,0)[b]{\hbox{\color{rgb_000000}\scriptsize $x_{1}$}}}
\put(1.2,1.37029){\makebox(0,0)[b]{\hbox{\color{rgb_000000}\scriptsize $x_{2}$}}}
\put(0.4,1.37029){\makebox(0,0)[b]{\hbox{\color{rgb_000000}\scriptsize $x_{3}$}}}
\put(1.8,0.229708){\makebox(0,0)[t]{\hbox{\color{rgb_000000}\scriptsize $y_{1}$}}}
\put(1.4,0.229708){\makebox(0,0)[t]{\hbox{\color{rgb_000000}\scriptsize $y_{2}$}}}
\put(0.2,0.229708){\makebox(0,0)[t]{\hbox{\color{rgb_000000}\scriptsize $y_{3}$}}}
\end{picture}%
%
    \caption{$\sigma=\tau_{1}$.}
    \label{figure:2b}
  \end{subfigure}
  \begin{subfigure}[t]{0.16\textwidth}
    \centering%
\xdefinecolor{rgb_000000}{rgb}{0,0,0}%
\xdefinecolor{rgb_0000ff}{rgb}{0,0,1}%
\xdefinecolor{rgb_e5e5e5}{rgb}{0.898039,0.898039,0.898039}%
\setlength{\unitlength}{1cm}%
\begin{picture}(2,1.6)(0,0)%
\color{rgb_e5e5e5}%
\path(0,0.3)(2,0.3)
\path(0,0.313889)(2,0.313889)
\path(0,0.327778)(2,0.327778)
\path(0,0.341667)(2,0.341667)
\path(0,0.355556)(2,0.355556)
\path(0,0.369444)(2,0.369444)
\path(0,0.383333)(2,0.383333)
\path(0,0.397222)(2,0.397222)
\path(0,0.411111)(2,0.411111)
\path(0,0.425)(2,0.425)
\path(0,0.438889)(2,0.438889)
\path(0,0.452778)(2,0.452778)
\path(0,0.466667)(2,0.466667)
\path(0,0.480556)(2,0.480556)
\path(0,0.494444)(2,0.494444)
\path(0,0.508333)(2,0.508333)
\path(0,0.522222)(2,0.522222)
\path(0,0.536111)(2,0.536111)
\path(0,0.55)(2,0.55)
\path(0,0.563889)(2,0.563889)
\path(0,0.577778)(2,0.577778)
\path(0,0.591667)(2,0.591667)
\path(0,0.605556)(2,0.605556)
\path(0,0.619444)(2,0.619444)
\path(0,0.633333)(2,0.633333)
\path(0,0.647222)(2,0.647222)
\path(0,0.661111)(2,0.661111)
\path(0,0.675)(2,0.675)
\path(0,0.688889)(2,0.688889)
\path(0,0.702778)(2,0.702778)
\path(0,0.716667)(2,0.716667)
\path(0,0.730556)(2,0.730556)
\path(0,0.744444)(2,0.744444)
\path(0,0.758333)(2,0.758333)
\path(0,0.772222)(2,0.772222)
\path(0,0.786111)(2,0.786111)
\path(0,0.8)(2,0.8)
\path(0,0.813889)(2,0.813889)
\path(0,0.827778)(2,0.827778)
\path(0,0.841667)(2,0.841667)
\path(0,0.855556)(2,0.855556)
\path(0,0.869444)(2,0.869444)
\path(0,0.883333)(2,0.883333)
\path(0,0.897222)(2,0.897222)
\path(0,0.911111)(2,0.911111)
\path(0,0.925)(2,0.925)
\path(0,0.938889)(2,0.938889)
\path(0,0.952778)(2,0.952778)
\path(0,0.966667)(2,0.966667)
\path(0,0.980556)(2,0.980556)
\path(0,0.994444)(2,0.994444)
\path(0,1.00833)(2,1.00833)
\path(0,1.02222)(2,1.02222)
\path(0,1.03611)(2,1.03611)
\path(0,1.05)(2,1.05)
\path(0,1.06389)(2,1.06389)
\path(0,1.07778)(2,1.07778)
\path(0,1.09167)(2,1.09167)
\path(0,1.10556)(2,1.10556)
\path(0,1.11944)(2,1.11944)
\path(0,1.13333)(2,1.13333)
\path(0,1.14722)(2,1.14722)
\path(0,1.16111)(2,1.16111)
\path(0,1.175)(2,1.175)
\path(0,1.18889)(2,1.18889)
\path(0,1.20278)(2,1.20278)
\path(0,1.21667)(2,1.21667)
\path(0,1.23056)(2,1.23056)
\path(0,1.24444)(2,1.24444)
\path(0,1.25833)(2,1.25833)
\path(0,1.27222)(2,1.27222)
\path(0,1.28611)(2,1.28611)
\path(0,1.3)(2,1.3)
\path(2,0.3)(2,1.3)
\path(1.98601,0.3)(1.98601,1.3)
\path(1.97203,0.3)(1.97203,1.3)
\path(1.95804,0.3)(1.95804,1.3)
\path(1.94406,0.3)(1.94406,1.3)
\path(1.93007,0.3)(1.93007,1.3)
\path(1.91608,0.3)(1.91608,1.3)
\path(1.9021,0.3)(1.9021,1.3)
\path(1.88811,0.3)(1.88811,1.3)
\path(1.87413,0.3)(1.87413,1.3)
\path(1.86014,0.3)(1.86014,1.3)
\path(1.84615,0.3)(1.84615,1.3)
\path(1.83217,0.3)(1.83217,1.3)
\path(1.81818,0.3)(1.81818,1.3)
\path(1.8042,0.3)(1.8042,1.3)
\path(1.79021,0.3)(1.79021,1.3)
\path(1.77622,0.3)(1.77622,1.3)
\path(1.76224,0.3)(1.76224,1.3)
\path(1.74825,0.3)(1.74825,1.3)
\path(1.73427,0.3)(1.73427,1.3)
\path(1.72028,0.3)(1.72028,1.3)
\path(1.70629,0.3)(1.70629,1.3)
\path(1.69231,0.3)(1.69231,1.3)
\path(1.67832,0.3)(1.67832,1.3)
\path(1.66434,0.3)(1.66434,1.3)
\path(1.65035,0.3)(1.65035,1.3)
\path(1.63636,0.3)(1.63636,1.3)
\path(1.62238,0.3)(1.62238,1.3)
\path(1.60839,0.3)(1.60839,1.3)
\path(1.59441,0.3)(1.59441,1.3)
\path(1.58042,0.3)(1.58042,1.3)
\path(1.56643,0.3)(1.56643,1.3)
\path(1.55245,0.3)(1.55245,1.3)
\path(1.53846,0.3)(1.53846,1.3)
\path(1.52448,0.3)(1.52448,1.3)
\path(1.51049,0.3)(1.51049,1.3)
\path(1.4965,0.3)(1.4965,1.3)
\path(1.48252,0.3)(1.48252,1.3)
\path(1.46853,0.3)(1.46853,1.3)
\path(1.45455,0.3)(1.45455,1.3)
\path(1.44056,0.3)(1.44056,1.3)
\path(1.42657,0.3)(1.42657,1.3)
\path(1.41259,0.3)(1.41259,1.3)
\path(1.3986,0.3)(1.3986,1.3)
\path(1.38462,0.3)(1.38462,1.3)
\path(1.37063,0.3)(1.37063,1.3)
\path(1.35664,0.3)(1.35664,1.3)
\path(1.34266,0.3)(1.34266,1.3)
\path(1.32867,0.3)(1.32867,1.3)
\path(1.31469,0.3)(1.31469,1.3)
\path(1.3007,0.3)(1.3007,1.3)
\path(1.28671,0.3)(1.28671,1.3)
\path(1.27273,0.3)(1.27273,1.3)
\path(1.25874,0.3)(1.25874,1.3)
\path(1.24476,0.3)(1.24476,1.3)
\path(1.23077,0.3)(1.23077,1.3)
\path(1.21678,0.3)(1.21678,1.3)
\path(1.2028,0.3)(1.2028,1.3)
\path(1.18881,0.3)(1.18881,1.3)
\path(1.17483,0.3)(1.17483,1.3)
\path(1.16084,0.3)(1.16084,1.3)
\path(1.14685,0.3)(1.14685,1.3)
\path(1.13287,0.3)(1.13287,1.3)
\path(1.11888,0.3)(1.11888,1.3)
\path(1.1049,0.3)(1.1049,1.3)
\path(1.09091,0.3)(1.09091,1.3)
\path(1.07692,0.3)(1.07692,1.3)
\path(1.06294,0.3)(1.06294,1.3)
\path(1.04895,0.3)(1.04895,1.3)
\path(1.03497,0.3)(1.03497,1.3)
\path(1.02098,0.3)(1.02098,1.3)
\path(1.00699,0.3)(1.00699,1.3)
\path(0.993007,0.3)(0.993007,1.3)
\path(0.979021,0.3)(0.979021,1.3)
\path(0.965035,0.3)(0.965035,1.3)
\path(0.951049,0.3)(0.951049,1.3)
\path(0.937063,0.3)(0.937063,1.3)
\path(0.923077,0.3)(0.923077,1.3)
\path(0.909091,0.3)(0.909091,1.3)
\path(0.895105,0.3)(0.895105,1.3)
\path(0.881119,0.3)(0.881119,1.3)
\path(0.867133,0.3)(0.867133,1.3)
\path(0.853147,0.3)(0.853147,1.3)
\path(0.839161,0.3)(0.839161,1.3)
\path(0.825175,0.3)(0.825175,1.3)
\path(0.811189,0.3)(0.811189,1.3)
\path(0.797203,0.3)(0.797203,1.3)
\path(0.783217,0.3)(0.783217,1.3)
\path(0.769231,0.3)(0.769231,1.3)
\path(0.755245,0.3)(0.755245,1.3)
\path(0.741259,0.3)(0.741259,1.3)
\path(0.727273,0.3)(0.727273,1.3)
\path(0.713287,0.3)(0.713287,1.3)
\path(0.699301,0.3)(0.699301,1.3)
\path(0.685315,0.3)(0.685315,1.3)
\path(0.671329,0.3)(0.671329,1.3)
\path(0.657343,0.3)(0.657343,1.3)
\path(0.643357,0.3)(0.643357,1.3)
\path(0.629371,0.3)(0.629371,1.3)
\path(0.615385,0.3)(0.615385,1.3)
\path(0.601399,0.3)(0.601399,1.3)
\path(0.587413,0.3)(0.587413,1.3)
\path(0.573427,0.3)(0.573427,1.3)
\path(0.559441,0.3)(0.559441,1.3)
\path(0.545455,0.3)(0.545455,1.3)
\path(0.531469,0.3)(0.531469,1.3)
\path(0.517483,0.3)(0.517483,1.3)
\path(0.503497,0.3)(0.503497,1.3)
\path(0.48951,0.3)(0.48951,1.3)
\path(0.475524,0.3)(0.475524,1.3)
\path(0.461538,0.3)(0.461538,1.3)
\path(0.447552,0.3)(0.447552,1.3)
\path(0.433566,0.3)(0.433566,1.3)
\path(0.41958,0.3)(0.41958,1.3)
\path(0.405594,0.3)(0.405594,1.3)
\path(0.391608,0.3)(0.391608,1.3)
\path(0.377622,0.3)(0.377622,1.3)
\path(0.363636,0.3)(0.363636,1.3)
\path(0.34965,0.3)(0.34965,1.3)
\path(0.335664,0.3)(0.335664,1.3)
\path(0.321678,0.3)(0.321678,1.3)
\path(0.307692,0.3)(0.307692,1.3)
\path(0.293706,0.3)(0.293706,1.3)
\path(0.27972,0.3)(0.27972,1.3)
\path(0.265734,0.3)(0.265734,1.3)
\path(0.251748,0.3)(0.251748,1.3)
\path(0.237762,0.3)(0.237762,1.3)
\path(0.223776,0.3)(0.223776,1.3)
\path(0.20979,0.3)(0.20979,1.3)
\path(0.195804,0.3)(0.195804,1.3)
\path(0.181818,0.3)(0.181818,1.3)
\path(0.167832,0.3)(0.167832,1.3)
\path(0.153846,0.3)(0.153846,1.3)
\path(0.13986,0.3)(0.13986,1.3)
\path(0.125874,0.3)(0.125874,1.3)
\path(0.111888,0.3)(0.111888,1.3)
\path(0.0979021,0.3)(0.0979021,1.3)
\path(0.0839161,0.3)(0.0839161,1.3)
\path(0.0699301,0.3)(0.0699301,1.3)
\path(0.0559441,0.3)(0.0559441,1.3)
\path(0.041958,0.3)(0.041958,1.3)
\path(0.027972,0.3)(0.027972,1.3)
\path(0.013986,0.3)(0.013986,1.3)
\path(0,0.3)(0,1.3)
\path(0,0.3)(2,0.3)(2,1.3)(0,1.3)(0,0.3)
\color{rgb_000000}%
\path(0,0.3)(2,0.3)
\path(0,1.3)(2,1.3)
\put(1.6,1.3){\color{rgb_000000}$\allinethickness{0.035146cm}\circle{0.035146}$}%
\put(1.2,1.3){\color{rgb_000000}$\allinethickness{0.035146cm}\circle{0.035146}$}%
\put(0.4,1.3){\color{rgb_000000}$\allinethickness{0.035146cm}\circle{0.035146}$}%
\put(1.8,0.3){\color{rgb_000000}$\allinethickness{0.035146cm}\circle{0.035146}$}%
\put(1.4,0.3){\color{rgb_000000}$\allinethickness{0.035146cm}\circle{0.035146}$}%
\put(0.2,0.3){\color{rgb_000000}$\allinethickness{0.035146cm}\circle{0.035146}$}%
\color{rgb_0000ff}%
\path(1.8,0.3)(1.6,1.3)
\path(1.4,0.3)(0.4,1.3)
\path(0.2,0.3)(1.2,1.3)
\put(1.6,1.37029){\makebox(0,0)[b]{\hbox{\color{rgb_000000}\scriptsize $x_{1}$}}}
\put(1.2,1.37029){\makebox(0,0)[b]{\hbox{\color{rgb_000000}\scriptsize $x_{2}$}}}
\put(0.4,1.37029){\makebox(0,0)[b]{\hbox{\color{rgb_000000}\scriptsize $x_{3}$}}}
\put(1.8,0.229708){\makebox(0,0)[t]{\hbox{\color{rgb_000000}\scriptsize $y_{1}$}}}
\put(1.4,0.229708){\makebox(0,0)[t]{\hbox{\color{rgb_000000}\scriptsize $y_{2}$}}}
\put(0.2,0.229708){\makebox(0,0)[t]{\hbox{\color{rgb_000000}\scriptsize $y_{3}$}}}
\end{picture}%
%
    \caption{$\sigma=\tau_{2}$.}
    \label{figure:2c}
  \end{subfigure}
  \begin{subfigure}[t]{0.16\textwidth}
    \centering%
\xdefinecolor{rgb_000000}{rgb}{0,0,0}%
\xdefinecolor{rgb_0000ff}{rgb}{0,0,1}%
\xdefinecolor{rgb_e5e5e5}{rgb}{0.898039,0.898039,0.898039}%
\setlength{\unitlength}{1cm}%
\begin{picture}(2,1.6)(0,0)%
\color{rgb_e5e5e5}%
\path(0,0.3)(2,0.3)
\path(0,0.313889)(2,0.313889)
\path(0,0.327778)(2,0.327778)
\path(0,0.341667)(2,0.341667)
\path(0,0.355556)(2,0.355556)
\path(0,0.369444)(2,0.369444)
\path(0,0.383333)(2,0.383333)
\path(0,0.397222)(2,0.397222)
\path(0,0.411111)(2,0.411111)
\path(0,0.425)(2,0.425)
\path(0,0.438889)(2,0.438889)
\path(0,0.452778)(2,0.452778)
\path(0,0.466667)(2,0.466667)
\path(0,0.480556)(2,0.480556)
\path(0,0.494444)(2,0.494444)
\path(0,0.508333)(2,0.508333)
\path(0,0.522222)(2,0.522222)
\path(0,0.536111)(2,0.536111)
\path(0,0.55)(2,0.55)
\path(0,0.563889)(2,0.563889)
\path(0,0.577778)(2,0.577778)
\path(0,0.591667)(2,0.591667)
\path(0,0.605556)(2,0.605556)
\path(0,0.619444)(2,0.619444)
\path(0,0.633333)(2,0.633333)
\path(0,0.647222)(2,0.647222)
\path(0,0.661111)(2,0.661111)
\path(0,0.675)(2,0.675)
\path(0,0.688889)(2,0.688889)
\path(0,0.702778)(2,0.702778)
\path(0,0.716667)(2,0.716667)
\path(0,0.730556)(2,0.730556)
\path(0,0.744444)(2,0.744444)
\path(0,0.758333)(2,0.758333)
\path(0,0.772222)(2,0.772222)
\path(0,0.786111)(2,0.786111)
\path(0,0.8)(2,0.8)
\path(0,0.813889)(2,0.813889)
\path(0,0.827778)(2,0.827778)
\path(0,0.841667)(2,0.841667)
\path(0,0.855556)(2,0.855556)
\path(0,0.869444)(2,0.869444)
\path(0,0.883333)(2,0.883333)
\path(0,0.897222)(2,0.897222)
\path(0,0.911111)(2,0.911111)
\path(0,0.925)(2,0.925)
\path(0,0.938889)(2,0.938889)
\path(0,0.952778)(2,0.952778)
\path(0,0.966667)(2,0.966667)
\path(0,0.980556)(2,0.980556)
\path(0,0.994444)(2,0.994444)
\path(0,1.00833)(2,1.00833)
\path(0,1.02222)(2,1.02222)
\path(0,1.03611)(2,1.03611)
\path(0,1.05)(2,1.05)
\path(0,1.06389)(2,1.06389)
\path(0,1.07778)(2,1.07778)
\path(0,1.09167)(2,1.09167)
\path(0,1.10556)(2,1.10556)
\path(0,1.11944)(2,1.11944)
\path(0,1.13333)(2,1.13333)
\path(0,1.14722)(2,1.14722)
\path(0,1.16111)(2,1.16111)
\path(0,1.175)(2,1.175)
\path(0,1.18889)(2,1.18889)
\path(0,1.20278)(2,1.20278)
\path(0,1.21667)(2,1.21667)
\path(0,1.23056)(2,1.23056)
\path(0,1.24444)(2,1.24444)
\path(0,1.25833)(2,1.25833)
\path(0,1.27222)(2,1.27222)
\path(0,1.28611)(2,1.28611)
\path(0,1.3)(2,1.3)
\path(2,0.3)(2,1.3)
\path(1.98601,0.3)(1.98601,1.3)
\path(1.97203,0.3)(1.97203,1.3)
\path(1.95804,0.3)(1.95804,1.3)
\path(1.94406,0.3)(1.94406,1.3)
\path(1.93007,0.3)(1.93007,1.3)
\path(1.91608,0.3)(1.91608,1.3)
\path(1.9021,0.3)(1.9021,1.3)
\path(1.88811,0.3)(1.88811,1.3)
\path(1.87413,0.3)(1.87413,1.3)
\path(1.86014,0.3)(1.86014,1.3)
\path(1.84615,0.3)(1.84615,1.3)
\path(1.83217,0.3)(1.83217,1.3)
\path(1.81818,0.3)(1.81818,1.3)
\path(1.8042,0.3)(1.8042,1.3)
\path(1.79021,0.3)(1.79021,1.3)
\path(1.77622,0.3)(1.77622,1.3)
\path(1.76224,0.3)(1.76224,1.3)
\path(1.74825,0.3)(1.74825,1.3)
\path(1.73427,0.3)(1.73427,1.3)
\path(1.72028,0.3)(1.72028,1.3)
\path(1.70629,0.3)(1.70629,1.3)
\path(1.69231,0.3)(1.69231,1.3)
\path(1.67832,0.3)(1.67832,1.3)
\path(1.66434,0.3)(1.66434,1.3)
\path(1.65035,0.3)(1.65035,1.3)
\path(1.63636,0.3)(1.63636,1.3)
\path(1.62238,0.3)(1.62238,1.3)
\path(1.60839,0.3)(1.60839,1.3)
\path(1.59441,0.3)(1.59441,1.3)
\path(1.58042,0.3)(1.58042,1.3)
\path(1.56643,0.3)(1.56643,1.3)
\path(1.55245,0.3)(1.55245,1.3)
\path(1.53846,0.3)(1.53846,1.3)
\path(1.52448,0.3)(1.52448,1.3)
\path(1.51049,0.3)(1.51049,1.3)
\path(1.4965,0.3)(1.4965,1.3)
\path(1.48252,0.3)(1.48252,1.3)
\path(1.46853,0.3)(1.46853,1.3)
\path(1.45455,0.3)(1.45455,1.3)
\path(1.44056,0.3)(1.44056,1.3)
\path(1.42657,0.3)(1.42657,1.3)
\path(1.41259,0.3)(1.41259,1.3)
\path(1.3986,0.3)(1.3986,1.3)
\path(1.38462,0.3)(1.38462,1.3)
\path(1.37063,0.3)(1.37063,1.3)
\path(1.35664,0.3)(1.35664,1.3)
\path(1.34266,0.3)(1.34266,1.3)
\path(1.32867,0.3)(1.32867,1.3)
\path(1.31469,0.3)(1.31469,1.3)
\path(1.3007,0.3)(1.3007,1.3)
\path(1.28671,0.3)(1.28671,1.3)
\path(1.27273,0.3)(1.27273,1.3)
\path(1.25874,0.3)(1.25874,1.3)
\path(1.24476,0.3)(1.24476,1.3)
\path(1.23077,0.3)(1.23077,1.3)
\path(1.21678,0.3)(1.21678,1.3)
\path(1.2028,0.3)(1.2028,1.3)
\path(1.18881,0.3)(1.18881,1.3)
\path(1.17483,0.3)(1.17483,1.3)
\path(1.16084,0.3)(1.16084,1.3)
\path(1.14685,0.3)(1.14685,1.3)
\path(1.13287,0.3)(1.13287,1.3)
\path(1.11888,0.3)(1.11888,1.3)
\path(1.1049,0.3)(1.1049,1.3)
\path(1.09091,0.3)(1.09091,1.3)
\path(1.07692,0.3)(1.07692,1.3)
\path(1.06294,0.3)(1.06294,1.3)
\path(1.04895,0.3)(1.04895,1.3)
\path(1.03497,0.3)(1.03497,1.3)
\path(1.02098,0.3)(1.02098,1.3)
\path(1.00699,0.3)(1.00699,1.3)
\path(0.993007,0.3)(0.993007,1.3)
\path(0.979021,0.3)(0.979021,1.3)
\path(0.965035,0.3)(0.965035,1.3)
\path(0.951049,0.3)(0.951049,1.3)
\path(0.937063,0.3)(0.937063,1.3)
\path(0.923077,0.3)(0.923077,1.3)
\path(0.909091,0.3)(0.909091,1.3)
\path(0.895105,0.3)(0.895105,1.3)
\path(0.881119,0.3)(0.881119,1.3)
\path(0.867133,0.3)(0.867133,1.3)
\path(0.853147,0.3)(0.853147,1.3)
\path(0.839161,0.3)(0.839161,1.3)
\path(0.825175,0.3)(0.825175,1.3)
\path(0.811189,0.3)(0.811189,1.3)
\path(0.797203,0.3)(0.797203,1.3)
\path(0.783217,0.3)(0.783217,1.3)
\path(0.769231,0.3)(0.769231,1.3)
\path(0.755245,0.3)(0.755245,1.3)
\path(0.741259,0.3)(0.741259,1.3)
\path(0.727273,0.3)(0.727273,1.3)
\path(0.713287,0.3)(0.713287,1.3)
\path(0.699301,0.3)(0.699301,1.3)
\path(0.685315,0.3)(0.685315,1.3)
\path(0.671329,0.3)(0.671329,1.3)
\path(0.657343,0.3)(0.657343,1.3)
\path(0.643357,0.3)(0.643357,1.3)
\path(0.629371,0.3)(0.629371,1.3)
\path(0.615385,0.3)(0.615385,1.3)
\path(0.601399,0.3)(0.601399,1.3)
\path(0.587413,0.3)(0.587413,1.3)
\path(0.573427,0.3)(0.573427,1.3)
\path(0.559441,0.3)(0.559441,1.3)
\path(0.545455,0.3)(0.545455,1.3)
\path(0.531469,0.3)(0.531469,1.3)
\path(0.517483,0.3)(0.517483,1.3)
\path(0.503497,0.3)(0.503497,1.3)
\path(0.48951,0.3)(0.48951,1.3)
\path(0.475524,0.3)(0.475524,1.3)
\path(0.461538,0.3)(0.461538,1.3)
\path(0.447552,0.3)(0.447552,1.3)
\path(0.433566,0.3)(0.433566,1.3)
\path(0.41958,0.3)(0.41958,1.3)
\path(0.405594,0.3)(0.405594,1.3)
\path(0.391608,0.3)(0.391608,1.3)
\path(0.377622,0.3)(0.377622,1.3)
\path(0.363636,0.3)(0.363636,1.3)
\path(0.34965,0.3)(0.34965,1.3)
\path(0.335664,0.3)(0.335664,1.3)
\path(0.321678,0.3)(0.321678,1.3)
\path(0.307692,0.3)(0.307692,1.3)
\path(0.293706,0.3)(0.293706,1.3)
\path(0.27972,0.3)(0.27972,1.3)
\path(0.265734,0.3)(0.265734,1.3)
\path(0.251748,0.3)(0.251748,1.3)
\path(0.237762,0.3)(0.237762,1.3)
\path(0.223776,0.3)(0.223776,1.3)
\path(0.20979,0.3)(0.20979,1.3)
\path(0.195804,0.3)(0.195804,1.3)
\path(0.181818,0.3)(0.181818,1.3)
\path(0.167832,0.3)(0.167832,1.3)
\path(0.153846,0.3)(0.153846,1.3)
\path(0.13986,0.3)(0.13986,1.3)
\path(0.125874,0.3)(0.125874,1.3)
\path(0.111888,0.3)(0.111888,1.3)
\path(0.0979021,0.3)(0.0979021,1.3)
\path(0.0839161,0.3)(0.0839161,1.3)
\path(0.0699301,0.3)(0.0699301,1.3)
\path(0.0559441,0.3)(0.0559441,1.3)
\path(0.041958,0.3)(0.041958,1.3)
\path(0.027972,0.3)(0.027972,1.3)
\path(0.013986,0.3)(0.013986,1.3)
\path(0,0.3)(0,1.3)
\path(0,0.3)(2,0.3)(2,1.3)(0,1.3)(0,0.3)
\color{rgb_000000}%
\path(0,0.3)(2,0.3)
\path(0,1.3)(2,1.3)
\put(1.6,1.3){\color{rgb_000000}$\allinethickness{0.035146cm}\circle{0.035146}$}%
\put(1.2,1.3){\color{rgb_000000}$\allinethickness{0.035146cm}\circle{0.035146}$}%
\put(0.4,1.3){\color{rgb_000000}$\allinethickness{0.035146cm}\circle{0.035146}$}%
\put(1.8,0.3){\color{rgb_000000}$\allinethickness{0.035146cm}\circle{0.035146}$}%
\put(1.4,0.3){\color{rgb_000000}$\allinethickness{0.035146cm}\circle{0.035146}$}%
\put(0.2,0.3){\color{rgb_000000}$\allinethickness{0.035146cm}\circle{0.035146}$}%
\color{rgb_0000ff}%
\path(1.8,0.3)(1.2,1.3)
\path(1.4,0.3)(0.4,1.3)
\path(0.2,0.3)(1.6,1.3)
\put(1.6,1.37029){\makebox(0,0)[b]{\hbox{\color{rgb_000000}\scriptsize $x_{1}$}}}
\put(1.2,1.37029){\makebox(0,0)[b]{\hbox{\color{rgb_000000}\scriptsize $x_{2}$}}}
\put(0.4,1.37029){\makebox(0,0)[b]{\hbox{\color{rgb_000000}\scriptsize $x_{3}$}}}
\put(1.8,0.229708){\makebox(0,0)[t]{\hbox{\color{rgb_000000}\scriptsize $y_{1}$}}}
\put(1.4,0.229708){\makebox(0,0)[t]{\hbox{\color{rgb_000000}\scriptsize $y_{2}$}}}
\put(0.2,0.229708){\makebox(0,0)[t]{\hbox{\color{rgb_000000}\scriptsize $y_{3}$}}}
\end{picture}%
%
    \caption{$\sigma=\tau_{1}\tau_{2}$.}
    \label{figure:2d}
  \end{subfigure}
  \begin{subfigure}[t]{0.16\textwidth}
    \centering%
\xdefinecolor{rgb_000000}{rgb}{0,0,0}%
\xdefinecolor{rgb_0000ff}{rgb}{0,0,1}%
\xdefinecolor{rgb_e5e5e5}{rgb}{0.898039,0.898039,0.898039}%
\setlength{\unitlength}{1cm}%
\begin{picture}(2,1.6)(0,0)%
\color{rgb_e5e5e5}%
\path(0,0.3)(2,0.3)
\path(0,0.313889)(2,0.313889)
\path(0,0.327778)(2,0.327778)
\path(0,0.341667)(2,0.341667)
\path(0,0.355556)(2,0.355556)
\path(0,0.369444)(2,0.369444)
\path(0,0.383333)(2,0.383333)
\path(0,0.397222)(2,0.397222)
\path(0,0.411111)(2,0.411111)
\path(0,0.425)(2,0.425)
\path(0,0.438889)(2,0.438889)
\path(0,0.452778)(2,0.452778)
\path(0,0.466667)(2,0.466667)
\path(0,0.480556)(2,0.480556)
\path(0,0.494444)(2,0.494444)
\path(0,0.508333)(2,0.508333)
\path(0,0.522222)(2,0.522222)
\path(0,0.536111)(2,0.536111)
\path(0,0.55)(2,0.55)
\path(0,0.563889)(2,0.563889)
\path(0,0.577778)(2,0.577778)
\path(0,0.591667)(2,0.591667)
\path(0,0.605556)(2,0.605556)
\path(0,0.619444)(2,0.619444)
\path(0,0.633333)(2,0.633333)
\path(0,0.647222)(2,0.647222)
\path(0,0.661111)(2,0.661111)
\path(0,0.675)(2,0.675)
\path(0,0.688889)(2,0.688889)
\path(0,0.702778)(2,0.702778)
\path(0,0.716667)(2,0.716667)
\path(0,0.730556)(2,0.730556)
\path(0,0.744444)(2,0.744444)
\path(0,0.758333)(2,0.758333)
\path(0,0.772222)(2,0.772222)
\path(0,0.786111)(2,0.786111)
\path(0,0.8)(2,0.8)
\path(0,0.813889)(2,0.813889)
\path(0,0.827778)(2,0.827778)
\path(0,0.841667)(2,0.841667)
\path(0,0.855556)(2,0.855556)
\path(0,0.869444)(2,0.869444)
\path(0,0.883333)(2,0.883333)
\path(0,0.897222)(2,0.897222)
\path(0,0.911111)(2,0.911111)
\path(0,0.925)(2,0.925)
\path(0,0.938889)(2,0.938889)
\path(0,0.952778)(2,0.952778)
\path(0,0.966667)(2,0.966667)
\path(0,0.980556)(2,0.980556)
\path(0,0.994444)(2,0.994444)
\path(0,1.00833)(2,1.00833)
\path(0,1.02222)(2,1.02222)
\path(0,1.03611)(2,1.03611)
\path(0,1.05)(2,1.05)
\path(0,1.06389)(2,1.06389)
\path(0,1.07778)(2,1.07778)
\path(0,1.09167)(2,1.09167)
\path(0,1.10556)(2,1.10556)
\path(0,1.11944)(2,1.11944)
\path(0,1.13333)(2,1.13333)
\path(0,1.14722)(2,1.14722)
\path(0,1.16111)(2,1.16111)
\path(0,1.175)(2,1.175)
\path(0,1.18889)(2,1.18889)
\path(0,1.20278)(2,1.20278)
\path(0,1.21667)(2,1.21667)
\path(0,1.23056)(2,1.23056)
\path(0,1.24444)(2,1.24444)
\path(0,1.25833)(2,1.25833)
\path(0,1.27222)(2,1.27222)
\path(0,1.28611)(2,1.28611)
\path(0,1.3)(2,1.3)
\path(2,0.3)(2,1.3)
\path(1.98601,0.3)(1.98601,1.3)
\path(1.97203,0.3)(1.97203,1.3)
\path(1.95804,0.3)(1.95804,1.3)
\path(1.94406,0.3)(1.94406,1.3)
\path(1.93007,0.3)(1.93007,1.3)
\path(1.91608,0.3)(1.91608,1.3)
\path(1.9021,0.3)(1.9021,1.3)
\path(1.88811,0.3)(1.88811,1.3)
\path(1.87413,0.3)(1.87413,1.3)
\path(1.86014,0.3)(1.86014,1.3)
\path(1.84615,0.3)(1.84615,1.3)
\path(1.83217,0.3)(1.83217,1.3)
\path(1.81818,0.3)(1.81818,1.3)
\path(1.8042,0.3)(1.8042,1.3)
\path(1.79021,0.3)(1.79021,1.3)
\path(1.77622,0.3)(1.77622,1.3)
\path(1.76224,0.3)(1.76224,1.3)
\path(1.74825,0.3)(1.74825,1.3)
\path(1.73427,0.3)(1.73427,1.3)
\path(1.72028,0.3)(1.72028,1.3)
\path(1.70629,0.3)(1.70629,1.3)
\path(1.69231,0.3)(1.69231,1.3)
\path(1.67832,0.3)(1.67832,1.3)
\path(1.66434,0.3)(1.66434,1.3)
\path(1.65035,0.3)(1.65035,1.3)
\path(1.63636,0.3)(1.63636,1.3)
\path(1.62238,0.3)(1.62238,1.3)
\path(1.60839,0.3)(1.60839,1.3)
\path(1.59441,0.3)(1.59441,1.3)
\path(1.58042,0.3)(1.58042,1.3)
\path(1.56643,0.3)(1.56643,1.3)
\path(1.55245,0.3)(1.55245,1.3)
\path(1.53846,0.3)(1.53846,1.3)
\path(1.52448,0.3)(1.52448,1.3)
\path(1.51049,0.3)(1.51049,1.3)
\path(1.4965,0.3)(1.4965,1.3)
\path(1.48252,0.3)(1.48252,1.3)
\path(1.46853,0.3)(1.46853,1.3)
\path(1.45455,0.3)(1.45455,1.3)
\path(1.44056,0.3)(1.44056,1.3)
\path(1.42657,0.3)(1.42657,1.3)
\path(1.41259,0.3)(1.41259,1.3)
\path(1.3986,0.3)(1.3986,1.3)
\path(1.38462,0.3)(1.38462,1.3)
\path(1.37063,0.3)(1.37063,1.3)
\path(1.35664,0.3)(1.35664,1.3)
\path(1.34266,0.3)(1.34266,1.3)
\path(1.32867,0.3)(1.32867,1.3)
\path(1.31469,0.3)(1.31469,1.3)
\path(1.3007,0.3)(1.3007,1.3)
\path(1.28671,0.3)(1.28671,1.3)
\path(1.27273,0.3)(1.27273,1.3)
\path(1.25874,0.3)(1.25874,1.3)
\path(1.24476,0.3)(1.24476,1.3)
\path(1.23077,0.3)(1.23077,1.3)
\path(1.21678,0.3)(1.21678,1.3)
\path(1.2028,0.3)(1.2028,1.3)
\path(1.18881,0.3)(1.18881,1.3)
\path(1.17483,0.3)(1.17483,1.3)
\path(1.16084,0.3)(1.16084,1.3)
\path(1.14685,0.3)(1.14685,1.3)
\path(1.13287,0.3)(1.13287,1.3)
\path(1.11888,0.3)(1.11888,1.3)
\path(1.1049,0.3)(1.1049,1.3)
\path(1.09091,0.3)(1.09091,1.3)
\path(1.07692,0.3)(1.07692,1.3)
\path(1.06294,0.3)(1.06294,1.3)
\path(1.04895,0.3)(1.04895,1.3)
\path(1.03497,0.3)(1.03497,1.3)
\path(1.02098,0.3)(1.02098,1.3)
\path(1.00699,0.3)(1.00699,1.3)
\path(0.993007,0.3)(0.993007,1.3)
\path(0.979021,0.3)(0.979021,1.3)
\path(0.965035,0.3)(0.965035,1.3)
\path(0.951049,0.3)(0.951049,1.3)
\path(0.937063,0.3)(0.937063,1.3)
\path(0.923077,0.3)(0.923077,1.3)
\path(0.909091,0.3)(0.909091,1.3)
\path(0.895105,0.3)(0.895105,1.3)
\path(0.881119,0.3)(0.881119,1.3)
\path(0.867133,0.3)(0.867133,1.3)
\path(0.853147,0.3)(0.853147,1.3)
\path(0.839161,0.3)(0.839161,1.3)
\path(0.825175,0.3)(0.825175,1.3)
\path(0.811189,0.3)(0.811189,1.3)
\path(0.797203,0.3)(0.797203,1.3)
\path(0.783217,0.3)(0.783217,1.3)
\path(0.769231,0.3)(0.769231,1.3)
\path(0.755245,0.3)(0.755245,1.3)
\path(0.741259,0.3)(0.741259,1.3)
\path(0.727273,0.3)(0.727273,1.3)
\path(0.713287,0.3)(0.713287,1.3)
\path(0.699301,0.3)(0.699301,1.3)
\path(0.685315,0.3)(0.685315,1.3)
\path(0.671329,0.3)(0.671329,1.3)
\path(0.657343,0.3)(0.657343,1.3)
\path(0.643357,0.3)(0.643357,1.3)
\path(0.629371,0.3)(0.629371,1.3)
\path(0.615385,0.3)(0.615385,1.3)
\path(0.601399,0.3)(0.601399,1.3)
\path(0.587413,0.3)(0.587413,1.3)
\path(0.573427,0.3)(0.573427,1.3)
\path(0.559441,0.3)(0.559441,1.3)
\path(0.545455,0.3)(0.545455,1.3)
\path(0.531469,0.3)(0.531469,1.3)
\path(0.517483,0.3)(0.517483,1.3)
\path(0.503497,0.3)(0.503497,1.3)
\path(0.48951,0.3)(0.48951,1.3)
\path(0.475524,0.3)(0.475524,1.3)
\path(0.461538,0.3)(0.461538,1.3)
\path(0.447552,0.3)(0.447552,1.3)
\path(0.433566,0.3)(0.433566,1.3)
\path(0.41958,0.3)(0.41958,1.3)
\path(0.405594,0.3)(0.405594,1.3)
\path(0.391608,0.3)(0.391608,1.3)
\path(0.377622,0.3)(0.377622,1.3)
\path(0.363636,0.3)(0.363636,1.3)
\path(0.34965,0.3)(0.34965,1.3)
\path(0.335664,0.3)(0.335664,1.3)
\path(0.321678,0.3)(0.321678,1.3)
\path(0.307692,0.3)(0.307692,1.3)
\path(0.293706,0.3)(0.293706,1.3)
\path(0.27972,0.3)(0.27972,1.3)
\path(0.265734,0.3)(0.265734,1.3)
\path(0.251748,0.3)(0.251748,1.3)
\path(0.237762,0.3)(0.237762,1.3)
\path(0.223776,0.3)(0.223776,1.3)
\path(0.20979,0.3)(0.20979,1.3)
\path(0.195804,0.3)(0.195804,1.3)
\path(0.181818,0.3)(0.181818,1.3)
\path(0.167832,0.3)(0.167832,1.3)
\path(0.153846,0.3)(0.153846,1.3)
\path(0.13986,0.3)(0.13986,1.3)
\path(0.125874,0.3)(0.125874,1.3)
\path(0.111888,0.3)(0.111888,1.3)
\path(0.0979021,0.3)(0.0979021,1.3)
\path(0.0839161,0.3)(0.0839161,1.3)
\path(0.0699301,0.3)(0.0699301,1.3)
\path(0.0559441,0.3)(0.0559441,1.3)
\path(0.041958,0.3)(0.041958,1.3)
\path(0.027972,0.3)(0.027972,1.3)
\path(0.013986,0.3)(0.013986,1.3)
\path(0,0.3)(0,1.3)
\path(0,0.3)(2,0.3)(2,1.3)(0,1.3)(0,0.3)
\color{rgb_000000}%
\path(0,0.3)(2,0.3)
\path(0,1.3)(2,1.3)
\put(1.6,1.3){\color{rgb_000000}$\allinethickness{0.035146cm}\circle{0.035146}$}%
\put(1.2,1.3){\color{rgb_000000}$\allinethickness{0.035146cm}\circle{0.035146}$}%
\put(0.4,1.3){\color{rgb_000000}$\allinethickness{0.035146cm}\circle{0.035146}$}%
\put(1.8,0.3){\color{rgb_000000}$\allinethickness{0.035146cm}\circle{0.035146}$}%
\put(1.4,0.3){\color{rgb_000000}$\allinethickness{0.035146cm}\circle{0.035146}$}%
\put(0.2,0.3){\color{rgb_000000}$\allinethickness{0.035146cm}\circle{0.035146}$}%
\color{rgb_0000ff}%
\path(1.8,0.3)(0.4,1.3)
\path(1.4,0.3)(1.6,1.3)
\path(0.2,0.3)(1.2,1.3)
\put(1.6,1.37029){\makebox(0,0)[b]{\hbox{\color{rgb_000000}\scriptsize $x_{1}$}}}
\put(1.2,1.37029){\makebox(0,0)[b]{\hbox{\color{rgb_000000}\scriptsize $x_{2}$}}}
\put(0.4,1.37029){\makebox(0,0)[b]{\hbox{\color{rgb_000000}\scriptsize $x_{3}$}}}
\put(1.8,0.229708){\makebox(0,0)[t]{\hbox{\color{rgb_000000}\scriptsize $y_{1}$}}}
\put(1.4,0.229708){\makebox(0,0)[t]{\hbox{\color{rgb_000000}\scriptsize $y_{2}$}}}
\put(0.2,0.229708){\makebox(0,0)[t]{\hbox{\color{rgb_000000}\scriptsize $y_{3}$}}}
\end{picture}%
%
    \caption{$\sigma=\tau_{2}\tau_{1}$.}
    \label{figure:2e}
  \end{subfigure}
  \begin{subfigure}[t]{0.16\textwidth}
    \centering%
\xdefinecolor{rgb_000000}{rgb}{0,0,0}%
\xdefinecolor{rgb_0000ff}{rgb}{0,0,1}%
\xdefinecolor{rgb_e5e5e5}{rgb}{0.898039,0.898039,0.898039}%
\setlength{\unitlength}{1cm}%
\begin{picture}(2,1.6)(0,0)%
\color{rgb_e5e5e5}%
\path(0,0.3)(2,0.3)
\path(0,0.313889)(2,0.313889)
\path(0,0.327778)(2,0.327778)
\path(0,0.341667)(2,0.341667)
\path(0,0.355556)(2,0.355556)
\path(0,0.369444)(2,0.369444)
\path(0,0.383333)(2,0.383333)
\path(0,0.397222)(2,0.397222)
\path(0,0.411111)(2,0.411111)
\path(0,0.425)(2,0.425)
\path(0,0.438889)(2,0.438889)
\path(0,0.452778)(2,0.452778)
\path(0,0.466667)(2,0.466667)
\path(0,0.480556)(2,0.480556)
\path(0,0.494444)(2,0.494444)
\path(0,0.508333)(2,0.508333)
\path(0,0.522222)(2,0.522222)
\path(0,0.536111)(2,0.536111)
\path(0,0.55)(2,0.55)
\path(0,0.563889)(2,0.563889)
\path(0,0.577778)(2,0.577778)
\path(0,0.591667)(2,0.591667)
\path(0,0.605556)(2,0.605556)
\path(0,0.619444)(2,0.619444)
\path(0,0.633333)(2,0.633333)
\path(0,0.647222)(2,0.647222)
\path(0,0.661111)(2,0.661111)
\path(0,0.675)(2,0.675)
\path(0,0.688889)(2,0.688889)
\path(0,0.702778)(2,0.702778)
\path(0,0.716667)(2,0.716667)
\path(0,0.730556)(2,0.730556)
\path(0,0.744444)(2,0.744444)
\path(0,0.758333)(2,0.758333)
\path(0,0.772222)(2,0.772222)
\path(0,0.786111)(2,0.786111)
\path(0,0.8)(2,0.8)
\path(0,0.813889)(2,0.813889)
\path(0,0.827778)(2,0.827778)
\path(0,0.841667)(2,0.841667)
\path(0,0.855556)(2,0.855556)
\path(0,0.869444)(2,0.869444)
\path(0,0.883333)(2,0.883333)
\path(0,0.897222)(2,0.897222)
\path(0,0.911111)(2,0.911111)
\path(0,0.925)(2,0.925)
\path(0,0.938889)(2,0.938889)
\path(0,0.952778)(2,0.952778)
\path(0,0.966667)(2,0.966667)
\path(0,0.980556)(2,0.980556)
\path(0,0.994444)(2,0.994444)
\path(0,1.00833)(2,1.00833)
\path(0,1.02222)(2,1.02222)
\path(0,1.03611)(2,1.03611)
\path(0,1.05)(2,1.05)
\path(0,1.06389)(2,1.06389)
\path(0,1.07778)(2,1.07778)
\path(0,1.09167)(2,1.09167)
\path(0,1.10556)(2,1.10556)
\path(0,1.11944)(2,1.11944)
\path(0,1.13333)(2,1.13333)
\path(0,1.14722)(2,1.14722)
\path(0,1.16111)(2,1.16111)
\path(0,1.175)(2,1.175)
\path(0,1.18889)(2,1.18889)
\path(0,1.20278)(2,1.20278)
\path(0,1.21667)(2,1.21667)
\path(0,1.23056)(2,1.23056)
\path(0,1.24444)(2,1.24444)
\path(0,1.25833)(2,1.25833)
\path(0,1.27222)(2,1.27222)
\path(0,1.28611)(2,1.28611)
\path(0,1.3)(2,1.3)
\path(2,0.3)(2,1.3)
\path(1.98601,0.3)(1.98601,1.3)
\path(1.97203,0.3)(1.97203,1.3)
\path(1.95804,0.3)(1.95804,1.3)
\path(1.94406,0.3)(1.94406,1.3)
\path(1.93007,0.3)(1.93007,1.3)
\path(1.91608,0.3)(1.91608,1.3)
\path(1.9021,0.3)(1.9021,1.3)
\path(1.88811,0.3)(1.88811,1.3)
\path(1.87413,0.3)(1.87413,1.3)
\path(1.86014,0.3)(1.86014,1.3)
\path(1.84615,0.3)(1.84615,1.3)
\path(1.83217,0.3)(1.83217,1.3)
\path(1.81818,0.3)(1.81818,1.3)
\path(1.8042,0.3)(1.8042,1.3)
\path(1.79021,0.3)(1.79021,1.3)
\path(1.77622,0.3)(1.77622,1.3)
\path(1.76224,0.3)(1.76224,1.3)
\path(1.74825,0.3)(1.74825,1.3)
\path(1.73427,0.3)(1.73427,1.3)
\path(1.72028,0.3)(1.72028,1.3)
\path(1.70629,0.3)(1.70629,1.3)
\path(1.69231,0.3)(1.69231,1.3)
\path(1.67832,0.3)(1.67832,1.3)
\path(1.66434,0.3)(1.66434,1.3)
\path(1.65035,0.3)(1.65035,1.3)
\path(1.63636,0.3)(1.63636,1.3)
\path(1.62238,0.3)(1.62238,1.3)
\path(1.60839,0.3)(1.60839,1.3)
\path(1.59441,0.3)(1.59441,1.3)
\path(1.58042,0.3)(1.58042,1.3)
\path(1.56643,0.3)(1.56643,1.3)
\path(1.55245,0.3)(1.55245,1.3)
\path(1.53846,0.3)(1.53846,1.3)
\path(1.52448,0.3)(1.52448,1.3)
\path(1.51049,0.3)(1.51049,1.3)
\path(1.4965,0.3)(1.4965,1.3)
\path(1.48252,0.3)(1.48252,1.3)
\path(1.46853,0.3)(1.46853,1.3)
\path(1.45455,0.3)(1.45455,1.3)
\path(1.44056,0.3)(1.44056,1.3)
\path(1.42657,0.3)(1.42657,1.3)
\path(1.41259,0.3)(1.41259,1.3)
\path(1.3986,0.3)(1.3986,1.3)
\path(1.38462,0.3)(1.38462,1.3)
\path(1.37063,0.3)(1.37063,1.3)
\path(1.35664,0.3)(1.35664,1.3)
\path(1.34266,0.3)(1.34266,1.3)
\path(1.32867,0.3)(1.32867,1.3)
\path(1.31469,0.3)(1.31469,1.3)
\path(1.3007,0.3)(1.3007,1.3)
\path(1.28671,0.3)(1.28671,1.3)
\path(1.27273,0.3)(1.27273,1.3)
\path(1.25874,0.3)(1.25874,1.3)
\path(1.24476,0.3)(1.24476,1.3)
\path(1.23077,0.3)(1.23077,1.3)
\path(1.21678,0.3)(1.21678,1.3)
\path(1.2028,0.3)(1.2028,1.3)
\path(1.18881,0.3)(1.18881,1.3)
\path(1.17483,0.3)(1.17483,1.3)
\path(1.16084,0.3)(1.16084,1.3)
\path(1.14685,0.3)(1.14685,1.3)
\path(1.13287,0.3)(1.13287,1.3)
\path(1.11888,0.3)(1.11888,1.3)
\path(1.1049,0.3)(1.1049,1.3)
\path(1.09091,0.3)(1.09091,1.3)
\path(1.07692,0.3)(1.07692,1.3)
\path(1.06294,0.3)(1.06294,1.3)
\path(1.04895,0.3)(1.04895,1.3)
\path(1.03497,0.3)(1.03497,1.3)
\path(1.02098,0.3)(1.02098,1.3)
\path(1.00699,0.3)(1.00699,1.3)
\path(0.993007,0.3)(0.993007,1.3)
\path(0.979021,0.3)(0.979021,1.3)
\path(0.965035,0.3)(0.965035,1.3)
\path(0.951049,0.3)(0.951049,1.3)
\path(0.937063,0.3)(0.937063,1.3)
\path(0.923077,0.3)(0.923077,1.3)
\path(0.909091,0.3)(0.909091,1.3)
\path(0.895105,0.3)(0.895105,1.3)
\path(0.881119,0.3)(0.881119,1.3)
\path(0.867133,0.3)(0.867133,1.3)
\path(0.853147,0.3)(0.853147,1.3)
\path(0.839161,0.3)(0.839161,1.3)
\path(0.825175,0.3)(0.825175,1.3)
\path(0.811189,0.3)(0.811189,1.3)
\path(0.797203,0.3)(0.797203,1.3)
\path(0.783217,0.3)(0.783217,1.3)
\path(0.769231,0.3)(0.769231,1.3)
\path(0.755245,0.3)(0.755245,1.3)
\path(0.741259,0.3)(0.741259,1.3)
\path(0.727273,0.3)(0.727273,1.3)
\path(0.713287,0.3)(0.713287,1.3)
\path(0.699301,0.3)(0.699301,1.3)
\path(0.685315,0.3)(0.685315,1.3)
\path(0.671329,0.3)(0.671329,1.3)
\path(0.657343,0.3)(0.657343,1.3)
\path(0.643357,0.3)(0.643357,1.3)
\path(0.629371,0.3)(0.629371,1.3)
\path(0.615385,0.3)(0.615385,1.3)
\path(0.601399,0.3)(0.601399,1.3)
\path(0.587413,0.3)(0.587413,1.3)
\path(0.573427,0.3)(0.573427,1.3)
\path(0.559441,0.3)(0.559441,1.3)
\path(0.545455,0.3)(0.545455,1.3)
\path(0.531469,0.3)(0.531469,1.3)
\path(0.517483,0.3)(0.517483,1.3)
\path(0.503497,0.3)(0.503497,1.3)
\path(0.48951,0.3)(0.48951,1.3)
\path(0.475524,0.3)(0.475524,1.3)
\path(0.461538,0.3)(0.461538,1.3)
\path(0.447552,0.3)(0.447552,1.3)
\path(0.433566,0.3)(0.433566,1.3)
\path(0.41958,0.3)(0.41958,1.3)
\path(0.405594,0.3)(0.405594,1.3)
\path(0.391608,0.3)(0.391608,1.3)
\path(0.377622,0.3)(0.377622,1.3)
\path(0.363636,0.3)(0.363636,1.3)
\path(0.34965,0.3)(0.34965,1.3)
\path(0.335664,0.3)(0.335664,1.3)
\path(0.321678,0.3)(0.321678,1.3)
\path(0.307692,0.3)(0.307692,1.3)
\path(0.293706,0.3)(0.293706,1.3)
\path(0.27972,0.3)(0.27972,1.3)
\path(0.265734,0.3)(0.265734,1.3)
\path(0.251748,0.3)(0.251748,1.3)
\path(0.237762,0.3)(0.237762,1.3)
\path(0.223776,0.3)(0.223776,1.3)
\path(0.20979,0.3)(0.20979,1.3)
\path(0.195804,0.3)(0.195804,1.3)
\path(0.181818,0.3)(0.181818,1.3)
\path(0.167832,0.3)(0.167832,1.3)
\path(0.153846,0.3)(0.153846,1.3)
\path(0.13986,0.3)(0.13986,1.3)
\path(0.125874,0.3)(0.125874,1.3)
\path(0.111888,0.3)(0.111888,1.3)
\path(0.0979021,0.3)(0.0979021,1.3)
\path(0.0839161,0.3)(0.0839161,1.3)
\path(0.0699301,0.3)(0.0699301,1.3)
\path(0.0559441,0.3)(0.0559441,1.3)
\path(0.041958,0.3)(0.041958,1.3)
\path(0.027972,0.3)(0.027972,1.3)
\path(0.013986,0.3)(0.013986,1.3)
\path(0,0.3)(0,1.3)
\path(0,0.3)(2,0.3)(2,1.3)(0,1.3)(0,0.3)
\color{rgb_000000}%
\path(0,0.3)(2,0.3)
\path(0,1.3)(2,1.3)
\put(1.6,1.3){\color{rgb_000000}$\allinethickness{0.035146cm}\circle{0.035146}$}%
\put(1.2,1.3){\color{rgb_000000}$\allinethickness{0.035146cm}\circle{0.035146}$}%
\put(0.4,1.3){\color{rgb_000000}$\allinethickness{0.035146cm}\circle{0.035146}$}%
\put(1.8,0.3){\color{rgb_000000}$\allinethickness{0.035146cm}\circle{0.035146}$}%
\put(1.4,0.3){\color{rgb_000000}$\allinethickness{0.035146cm}\circle{0.035146}$}%
\put(0.2,0.3){\color{rgb_000000}$\allinethickness{0.035146cm}\circle{0.035146}$}%
\color{rgb_0000ff}%
\path(1.8,0.3)(0.4,1.3)
\path(1.4,0.3)(1.2,1.3)
\path(0.2,0.3)(1.6,1.3)
\put(1.6,1.37029){\makebox(0,0)[b]{\hbox{\color{rgb_000000}\scriptsize $x_{1}$}}}
\put(1.2,1.37029){\makebox(0,0)[b]{\hbox{\color{rgb_000000}\scriptsize $x_{2}$}}}
\put(0.4,1.37029){\makebox(0,0)[b]{\hbox{\color{rgb_000000}\scriptsize $x_{3}$}}}
\put(1.8,0.229708){\makebox(0,0)[t]{\hbox{\color{rgb_000000}\scriptsize $y_{1}$}}}
\put(1.4,0.229708){\makebox(0,0)[t]{\hbox{\color{rgb_000000}\scriptsize $y_{2}$}}}
\put(0.2,0.229708){\makebox(0,0)[t]{\hbox{\color{rgb_000000}\scriptsize $y_{3}$}}}
\end{picture}%
%
    \caption{$\sigma=\tau_{1}\tau_{2}\tau_{1}$.}
    \label{figure:2f}
  \end{subfigure}
  \caption{Typical trajectories contained in the weighted sum
    $K_{\mathcal{M}_{3}}(\bm{x},\bm{y};t)=\sum_{\sigma\in
      S_{3}}\chi(\sigma)K_{\mathring{\mathbb{R}}^{3}}(\bm{x},\sigma\bm{y};t)$. Any
    permutation $\sigma\in S_{3}$ can be decomposed into a product of
    the adjacent transpositions $\tau_{1}=(1,2)$ and
    $\tau_{2}=(2,3)$. Correspondingly, any weight factor
    $\chi(\sigma)$ can be decomposed into a product of
    $\chi(\tau_{1})$ and $\chi(\tau_{2})$. The physical meaning of
    this is that the Feynman kernel acquires the particle-exchange
    phase $\chi(\tau_{j})$ every time two adjacent particles collide
    with each other in the course of the time-evolution. This is
    equivalent to the statement that the Feynman kernel acquires the
    particle-exchange phase $\chi(\tau_{j})$ every time the
    three-particle trajectory hits the codimension-1 boundary
    $\partial\mathcal{M}_{3,j}^{\text{2-body}}$.}
  \label{figure:2}
\end{figure}

Let $\psi(\bm{x},t)$ be an equivariant function on
$\mathring{\mathbb{R}}^{n}$ that satisfies
$\psi(\sigma\bm{x},t)=\chi(\sigma)\psi(\bm{x},t)$ for any
$\sigma\in S_{n}$. Notice that, if $\chi=\chi^{\text{[B]}}$
($\chi=\chi^{\text{[F]}}$), such an equivariant function is nothing
but the wavefunction of identical bosons (fermions) on
$\mathring{\mathbb{R}}^{n}$. Then, the solution to the time-dependent
Schr\"{o}dinger equation on $\mathring{\mathbb{R}}^{n}$ is given by
the following integral transform of the initial wavefunction:
\begin{align}
  \psi(\bm{x},t)
  &=\int_{\mathring{\mathbb{R}}^{n}}\!\!d\bm{y}\,K_{\mathring{\mathbb{R}}^{n}}(\bm{x},\bm{y};t)\psi(\bm{y},0)\nonumber\\
  &=\int_{\mathcal{M}_{n}}\!\!d\bm{y}\left(\sum_{\sigma\in S_{n}}K_{\mathring{\mathbb{R}}^{n}}(\bm{x},\sigma\bm{y};t)\psi(\sigma\bm{y},0)\right)\nonumber\\
  &=\int_{\mathcal{M}_{n}}\!\!d\bm{y}\left(\sum_{\sigma\in S_{n}}K_{\mathring{\mathbb{R}}^{n}}(\bm{x},\sigma\bm{y};t)\chi(\sigma)\psi(\bm{y},0)\right)\nonumber\\
  &=\int_{\mathcal{M}_{n}}\!\!d\bm{y}\left(\sum_{\sigma\in S_{n}}\chi(\sigma)K_{\mathring{\mathbb{R}}^{n}}(\bm{x},\sigma\bm{y};t)\right)\psi(\bm{y},0),\quad\forall\bm{x}\in\mathring{\mathbb{R}}^{n}.\label{eq:39}
\end{align}
Here the second equality follows from the following integral formula
(for the proof, see appendix \ref{appendix:B}):
\begin{align}
  \int_{\mathring{\mathbb{R}}^{n}}\!\!d\bm{y}\,f(\bm{y})=\int_{\mathcal{M}_{n}}\!\!d\bm{y}\left(\sum_{\sigma\in S_{n}}f(\sigma\bm{y})\right),\label{eq:40}
\end{align}
where $f$ is an arbitrary test function on
$\mathring{\mathbb{R}}^{n}$. By restricting the variable $\bm{x}$ to
$\mathcal{M}_{n}$ and then comparing \eqref{eq:39} with \eqref{eq:32},
one arrives at the formula \eqref{eq:33}. A proof that
eq.~\eqref{eq:33} indeed satisfies the conditions
\eqref{eq:27}--\eqref{eq:30} is presented in appendix
\ref{appendix:A}. It is now obvious from the above derivation that the
weight factor $\chi(\sigma)$ in \eqref{eq:33} describes a
particle-exchange phase. More precisely, $\chi(\sigma)$ is an
accumulation of particle-exchange phases for two adjacent particles in
the course of the time-evolution; see figure \ref{figure:2} for the
case of $n=3$.

Now, as noted repeatedly, there are just two distinct one-dimensional
unitary representations of $S_{n}$: the totally symmetric
representation $\chi^{\text{[B]}}$ and the totally antisymmetric
representation $\chi^{\text{[F]}}$. This simple mathematical fact has
two distinct physical meanings here. To see this, let us first suppose
that $K_{\mathring{\mathbb{R}}^{n}}$ is given. Then the formula
\eqref{eq:33} implies that there exist two distinct Feynman kernels on
$\mathcal{M}_{n}$:
\begin{subequations}
  \begin{align}
    K^{\text{[B]}}_{\mathcal{M}_{n}}(\bm{x},\bm{y};t)&=\sum_{\sigma\in S_{n}}\chi^{\text{[B]}}(\sigma)K_{\mathring{\mathbb{R}}^{n}}(\bm{x},\sigma\bm{y};t),\label{eq:41a}\\
    K^{\text{[F]}}_{\mathcal{M}_{n}}(\bm{x},\bm{y};t)&=\sum_{\sigma\in S_{n}}\chi^{\text{[F]}}(\sigma)K_{\mathring{\mathbb{R}}^{n}}(\bm{x},\sigma\bm{y};t).\label{eq:41b}
  \end{align}
\end{subequations}
Since $K_{\mathring{\mathbb{R}}^{n}}$ is normally constructed from
classical theory via the Feynman's path-integral quantization,
eqs.~\eqref{eq:41a} and \eqref{eq:41b} indicate that there exist two
inequivalent quantizations depending on the particle-exchange phases
$\chi^{\text{[B/F]}}(\sigma)$. Hence the first meaning is about the
particle statistics: in one dimension, there just exists the
Bose-Fermi alternative rather than the intermediate statistics of
Leinaas and Myrheim \cite{Leinaas:1977fm}. To see another meaning, let
us next suppose that $K_{\mathcal{M}_{n}}$ is given. Then the formula
\eqref{eq:33} implies that there may exist two distinct Feynman
kernels on $\mathring{\mathbb{R}}^{n}$ satisfying the following
equalities:
\begin{align}
  K_{\mathcal{M}_{n}}(\bm{x},\bm{y};t)
  &=\sum_{\sigma\in S_{n}}\chi^{\text{[B]}}(\sigma)K^{\text{[B]}}_{\mathring{\mathbb{R}}^{n}}(\bm{x},\sigma\bm{y};t)\nonumber\\
  &=\sum_{\sigma\in S_{n}}\chi^{\text{[F]}}(\sigma)K^{\text{[F]}}_{\mathring{\mathbb{R}}^{n}}(\bm{x},\sigma\bm{y};t).\label{eq:42}
\end{align}
Since $K_{\mathcal{M}_{n}}$ determines the dynamics as well as energy
spectrum of identical-particle systems, eq.~\eqref{eq:42} indicates
that the two distinct systems on $\mathring{\mathbb{R}}^{n}$ described
by $K^{\text{[B]}}_{\mathring{\mathbb{R}}^{n}}$ and
$K^{\text{[F]}}_{\mathring{\mathbb{R}}^{n}}$ are, if they exist,
completely isospectral. Hence the second meaning is about the
boson-fermion duality: there may exist bosonic and fermionic systems
whose energy spectra are completely equivalent.

Let us finally derive a generalization of the boson-fermion duality by
using \eqref{eq:42}, which can be achieved by studying the connection
conditions for $K^{\text{[B]}}_{\mathring{\mathbb{R}}^{n}}$ and
$K^{\text{[F]}}_{\mathring{\mathbb{R}}^{n}}$ at the codimension-1
singularities. To this end, let us first note that any element of
$S_{n}$ can be classified into either even or odd permutations. In
other words, the symmetric group $S_{n}$ can be decomposed into the
following coset decomposition:
\begin{align}
  S_{n}=A_{n}\cup A_{n}\tau,\label{eq:43}
\end{align}
where $A_{n}$ is the alternating group that consists of only even
permutations. $\tau\in S_{n}$ is an arbitrary transposition and
$A_{n}\tau=\{\sigma\tau:\sigma\in A_{n}\}$ is the right coset that
consists of only odd permutations. Then, corresponding to
\eqref{eq:43}, the formula \eqref{eq:33} can be decomposed into the
following form:
\begin{align}
  K_{\mathcal{M}_{n}}(\bm{x},\bm{y};t)
  &=\sum_{\sigma\in A_{n}}\left(\chi(\sigma)K_{\mathring{\mathbb{R}}^{n}}(\bm{x},\sigma\bm{y};t)+\chi(\sigma\tau)K_{\mathring{\mathbb{R}}^{n}}(\bm{x},\sigma\tau\bm{y};t)\right)\nonumber\\
  &=\sum_{\sigma\in A_{n}}\chi(\sigma)\left(K_{\mathring{\mathbb{R}}^{n}}(\sigma^{-1}\bm{x},\bm{y};t)+\chi(\tau)K_{\mathring{\mathbb{R}}^{n}}(\tau^{-1}\sigma^{-1}\bm{x},\bm{y};t)\right),\label{eq:44}
\end{align}
where we have used $\chi(\sigma\tau)=\chi(\sigma)\chi(\tau)$ and
$K_{\mathring{\mathbb{R}}^{n}}(\bm{x},\sigma\bm{y};t)=K_{\mathring{\mathbb{R}}^{n}}(\sigma\sigma^{-1}\bm{x},\sigma\bm{y};t)=K_{\mathring{\mathbb{R}}^{n}}(\sigma^{-1}\bm{x},\bm{y};t)$
($\forall\sigma\in S_{n}$) which follows from the permutation
invariance \eqref{eq:38}. Now we choose $\tau$ as the adjacent
transposition $\tau=\tau_{j}=(j,j+1)$ which just swaps $x_{j}$ and
$x_{j+1}$. Then, by applying the differential operator
$\frac{\partial}{\partial x_{j}}-\frac{\partial}{\partial
  x_{j+1}}-\frac{1}{a_{j}}$ to \eqref{eq:44}, we get
\begin{align}
  &\left(\frac{\partial}{\partial x_{j}}-\frac{\partial}{\partial x_{j+1}}\right)K_{\mathcal{M}_{n}}(\bm{x},\bm{y};t)-\frac{1}{a_{j}}K_{\mathcal{M}_{n}}(\bm{x},\bm{y};t)\nonumber\\
  &=\sum_{\sigma\in A_{n}}\chi(\sigma)\biggl[\left(\frac{\partial}{\partial x_{j}}-\frac{\partial}{\partial x_{j+1}}\right)K_{\mathring{\mathbb{R}}^{n}}(\sigma^{-1}\bm{x},\bm{y};t)\nonumber\\
  &\quad
    -\chi(\tau_{j})\left(\frac{\partial}{\partial x_{j+1}}-\frac{\partial}{\partial x_{j}}\right)K_{\mathring{\mathbb{R}}^{n}}(\tau_{j}^{-1}\sigma^{-1}\bm{x},\bm{y};t)\nonumber\\
  &\quad
    -\frac{1}{a_{j}}\biggl(K_{\mathring{\mathbb{R}}^{n}}(\sigma^{-1}\bm{x},\bm{y};t)+\chi(\tau_{j})K_{\mathring{\mathbb{R}}^{n}}(\tau_{j}^{-1}\sigma^{-1}\bm{x},\bm{y};t)\biggr)\biggr]\nonumber\\
  &=\sum_{\sigma\in A_{n}}\chi(\sigma)\biggl[\left(\frac{\partial}{\partial z_{\sigma(j)}}-\frac{\partial}{\partial z_{\sigma(j+1)}}\right)K_{\mathring{\mathbb{R}}^{n}}(\bm{z},\bm{y};t)\biggr|_{\bm{z}=\sigma^{-1}\bm{x}}\nonumber\\
  &\quad
    -\chi(\tau_{j})\left(\frac{\partial}{\partial z_{\sigma(j)}}-\frac{\partial}{\partial z_{\sigma(j+1)}}\right)K_{\mathring{\mathbb{R}}^{n}}(\bm{z},\bm{y};t)\biggr|_{\bm{z}=\tau_{j}^{-1}\sigma^{-1}\bm{x}}\nonumber\\
  &\quad
    -\frac{1}{a_{j}}\biggl(K_{\mathring{\mathbb{R}}^{n}}(\bm{z},\bm{y};t)\biggr|_{\bm{z}=\sigma^{-1}\bm{x}}+\chi(\tau_{j})K_{\mathring{\mathbb{R}}^{n}}(\bm{z},\bm{y};t)\biggr|_{\bm{z}=\tau_{j}^{-1}\sigma^{-1}\bm{x}}\biggr)\biggr],\label{eq:45}
\end{align}
where in the last equality we have introduced a new variable
$\bm{z}=\sigma^{-1}\bm{x}$ and used the relation $\bm{x}=\sigma\bm{z}$
(i.e., $x_{k}=z_{\sigma(k)}$ for any $k$) in the first and third
terms. Similarly, in the second and fourth terms of the last line we
have introduced $\bm{z}=\tau_{j}^{-1}\sigma^{-1}\bm{x}$ and used the
relation $\bm{x}=\sigma\tau_{j}\bm{z}$, which gives
$x_{j+1}=z_{\sigma(\tau_{j}(j+1))}=z_{\sigma(j)}$,
$x_{j}=z_{\sigma(\tau_{j}(j))}=z_{\sigma(j+1)}$, and
$x_{k}=z_{\sigma(\tau_{j}(k))}=z_{\sigma(k)}$ for
$k\notin\{j,j+1\}$. It should be noted that $\bm{x}$ in \eqref{eq:45}
is in the region $x_{1}>\cdots>x_{n}$, which means that $\bm{z}$ in
the first and third terms is in the region
$z_{\sigma(1)}>\cdots>z_{\sigma(j)}>z_{\sigma(j+1)}>\cdots>z_{\sigma(n)}$
whereas $\bm{z}$ in the second and fourth terms is in the region
$z_{\sigma(1)}>\cdots>z_{\sigma(j+1)}>z_{\sigma(j)}>\cdots>z_{\sigma(n)}$. Hence,
in order for $K_{\mathcal{M}_{n}}$ to satisfy the Robin boundary
condition \eqref{eq:31} as $x_{j}-x_{j+1}\to0_{+}$, the Feynman kernel
on $\mathring{\mathbb{R}}^{n}$ should satisfy the following connection
condition at the codimension-1 singularity
$\{z_{\sigma(1)}>\cdots>z_{\sigma(j)}=z_{\sigma(j+1)}>\cdots>z_{\sigma(n)}\}$:
\begin{align}
  0
  &=\left(\frac{\partial}{\partial z_{\sigma(j)}}-\frac{\partial}{\partial z_{\sigma(j+1)}}\right)K_{\mathring{\mathbb{R}}^{n}}(\bm{z},\bm{y};t)\biggr|_{z_{\sigma(j)}-z_{\sigma(j+1)}=0_{+}}\nonumber\\
  &\quad
    -\chi(\tau_{j})\left(\frac{\partial}{\partial z_{\sigma(j)}}-\frac{\partial}{\partial z_{\sigma(j+1)}}\right)K_{\mathring{\mathbb{R}}^{n}}(\bm{z},\bm{y};t)\biggr|_{z_{\sigma(j+1)}-z_{\sigma(j)}=0_{+}}\nonumber\\
  &\quad
    -\frac{1}{a_{j}}\biggl(K_{\mathring{\mathbb{R}}^{n}}(\bm{z},\bm{y};t)\biggr|_{z_{\sigma(j)}-z_{\sigma(j+1)}=0_{+}}
    +\chi(\tau_{j})K_{\mathring{\mathbb{R}}^{n}}(\bm{z},\bm{y};t)\biggr|_{z_{\sigma(j+1)}-z_{\sigma(j)}=0_{+}}\biggr),\label{eq:46}
\end{align}
where $j\in\{1,\cdots,n-1\}$ and $\sigma\in A_{n}$. Now it is obvious
that, in the totally symmetric representation
$\chi=\chi^{\text{[B]}}$, where $\chi^{\text{[B]}}(\tau_{j})=1$, we
have the following connection condition for
$K_{\mathring{\mathbb{R}}^{n}}=K_{\mathring{\mathbb{R}}^{n}}^{\text{[B]}}$:
\begin{align}
  0
  &=\left(\frac{\partial}{\partial x_{\sigma(j)}}-\frac{\partial}{\partial x_{\sigma(j+1)}}\right)K^{\text{[B]}}_{\mathring{\mathbb{R}}^{n}}(\bm{x},\bm{y};t)\biggr|_{x_{\sigma(j)}-x_{\sigma(j+1)}=0_{+}}\nonumber\\
  &\quad
    -\left(\frac{\partial}{\partial x_{\sigma(j)}}-\frac{\partial}{\partial x_{\sigma(j+1)}}\right)K^{\text{[B]}}_{\mathring{\mathbb{R}}^{n}}(\bm{x},\bm{y};t)\biggr|_{x_{\sigma(j)}-x_{\sigma(j+1)}=0_{-}}\nonumber\\
  &\quad
    -\frac{1}{a_{j}}\left(K^{\text{[B]}}_{\mathring{\mathbb{R}}^{n}}(\bm{x},\bm{y};t)\biggr|_{x_{\sigma(j)}-x_{\sigma(j+1)}=0_{+}}
    +K^{\text{[B]}}_{\mathring{\mathbb{R}}^{n}}(\bm{x},\bm{y};t)\biggr|_{x_{\sigma(j)}-x_{\sigma(j+1)}=0_{-}}\right),\label{eq:47}
\end{align}
where we have renamed $\bm{z}$ to $\bm{x}$. In contrast, in the
totally antisymmetric representation $\chi=\chi^{\text{[F]}}$, where
$\chi^{\text{[F]}}(\tau_{j})=\operatorname{sgn}(\tau_{j})=-1$, we have
the following connection condition for
$K_{\mathring{\mathbb{R}}^{n}}=K_{\mathring{\mathbb{R}}^{n}}^{\text{[F]}}$:
\begin{align}
  0
  &=\left(\frac{\partial}{\partial x_{\sigma(j)}}-\frac{\partial}{\partial x_{\sigma(j+1)}}\right)K^{\text{[F]}}_{\mathring{\mathbb{R}}^{n}}(\bm{x},\bm{y};t)\biggr|_{x_{\sigma(j)}-x_{\sigma(j+1)}=0_{+}}\nonumber\\
  &\quad
    +\left(\frac{\partial}{\partial x_{\sigma(j)}}-\frac{\partial}{\partial x_{\sigma(j+1)}}\right)K^{\text{[F]}}_{\mathring{\mathbb{R}}^{n}}(\bm{x},\bm{y};t)\biggr|_{x_{\sigma(j)}-x_{\sigma(j+1)}=0_{-}}\nonumber\\
  &\quad
    -\frac{1}{a_{j}}\left(K^{\text{[F]}}_{\mathring{\mathbb{R}}^{n}}(\bm{x},\bm{y};t)\biggr|_{x_{\sigma(j)}-x_{\sigma(j+1)}=0_{+}}
    -K^{\text{[F]}}_{\mathring{\mathbb{R}}^{n}}(\bm{x},\bm{y};t)\biggr|_{x_{\sigma(j)}-x_{\sigma(j+1)}=0_{-}}\right).\label{eq:48}
\end{align}

Now, these connection conditions are nothing but (a part of) the
connection conditions for the $\delta$- and $\varepsilon$-function
potentials. In fact, it follows from these connection conditions that
the wavefunctions constructed from the integral transform
\eqref{eq:39} satisfy the conditions \eqref{eq:2a} or \eqref{eq:3a}
with $a$ being replaced by $a_{j}$. Note that the continuity
conditions for the wavefunction \eqref{eq:2b} and its derivative
\eqref{eq:3b} are not necessary here because the totally symmetric
function and the derivative of totally antisymmetric function
automatically becomes continuous at the coincidence points. Thus,
reversing the argument used to arrive at \eqref{eq:2a}--\eqref{eq:3b}
from \eqref{eq:1a} and \eqref{eq:1b}, we obtain the following dual
Hamiltonians for $n$ identical bosons and fermions:
\begin{subequations}
  \begin{align}
    H_{\text{B}}&=-\frac{\hbar^{2}}{2m}\bm{\nabla}^{2}+V_{\text{B}}(\bm{x}),\label{eq:49a}\\
    H_{\text{F}}&=-\frac{\hbar^{2}}{2m}\bm{\nabla}^{2}+V_{\text{F}}(\bm{x}),\label{eq:49b}
  \end{align}
\end{subequations}
where
\begin{subequations}
  \begin{align}
    V_{\text{B}}(\bm{x})&=\frac{\hbar^{2}}{m}\sum_{j=1}^{n-1}\sum_{\sigma\in A_{n}}\left[\prod_{k\in\{1,\cdots,n-1\}\setminus\{j\}}\theta(x_{\sigma(k)}-x_{\sigma(k+1)})\right]\delta(x_{\sigma(j)}-x_{\sigma(j+1)};\tfrac{1}{a_{j}}),\label{eq:50a}\\
    V_{\text{F}}(\bm{x})&=\frac{\hbar^{2}}{m}\sum_{j=1}^{n-1}\sum_{\sigma\in A_{n}}\left[\prod_{k\in\{1,\cdots,n-1\}\setminus\{j\}}\theta(x_{\sigma(k)}-x_{\sigma(k+1)})\right]\varepsilon(x_{\sigma(j)}-x_{\sigma(j+1)};a_{j}).\label{eq:50b}
  \end{align}
\end{subequations}
Note that the factor
$\prod_{k\in\{1,\cdots,n-1\}\setminus\{j\}}\theta(x_{\sigma(k)}-x_{\sigma(k+1)})$,
where $\theta$ is the step function, is introduced in order to
guarantee the ordering
$x_{\sigma(1)}>\cdots>x_{\sigma(j)}=x_{\sigma(j+1)}>\cdots>x_{\sigma(n)}$. Note
also that the total number of the codimension-1 singularities in
$\mathring{\mathbb{R}}^{n}$ is
$\frac{n(n-1)}{2}\times(n-1)!=(n-1)\times\frac{n!}{2}=(n-1)\times|A_{n}|$,
where $|A_{n}|$ stands for the order of the alternating group
$A_{n}$. Hence the summation $\sum_{j=1}^{n-1}\sum_{\sigma\in A_{n}}$
is indeed summing over all the codimension-1 singularities. Finally,
it is easy to see that the potential energies \eqref{eq:50a} and
\eqref{eq:50b} are invariant under $S_{n}$ and satisfy the identities
$V_{\text{B/F}}(\sigma\bm{x})=V_{\text{B/F}}(\bm{x})$ for any
$\sigma\in S_{n}$.

To summarize, by imposing the Robin boundary conditions to the Feynman
kernel \eqref{eq:42}, we have found that
$K_{\mathring{\mathbb{R}}^{n}}^{\text{[B]}}$ and
$K_{\mathring{\mathbb{R}}^{n}}^{\text{[F]}}$, respectively, must
satisfy the connection conditions for the $\delta$- and
$\varepsilon$-function potentials at the codimension-1 singularities
in $\mathring{\mathbb{R}}^{n}$. The $n$-body Hamiltonians that realize
these connection conditions for identical bosons and fermions are
given by \eqref{eq:49a} and \eqref{eq:49b}, both of which possess
$n-1$ distinct (coordinate-dependent) coupling constants. By
construction, these models enjoy (i) the spectral equivalence, (ii)
the boson-fermion mapping, and (iii) the strong-weak duality.

\section{Summary and discussion}
\label{section:4}
In the present paper, we have revisited the boson-fermion duality in
one dimension by using the configuration-space approach and the
path-integral formalism. In section \ref{section:1}, we have first
presented the detailed analysis of the configuration space for $n$
identical particles on $\mathbb{R}$. We have shown that the two-body
contact interactions for $n$ identical particles are generally
described by the $n-1$ distinct Robin boundary conditions
\eqref{eq:23}, where the boundary-condition parameters $a_{j}$
($j=1,\cdots,n-1$) may depend on the coordinates parallel to the
codimension-1 boundaries of
$\mathcal{M}_{n}=\mathring{\mathbb{R}}^{n}/S_{n}$. In section
\ref{section:2}, we have then studied the dynamics of $n$ identical
particles on $\mathbb{R}$ by using the Feynman kernel. We have first
shown that the Feynman kernel on $\mathcal{M}_{n}$ is generally given
by \eqref{eq:33}---the weighted sum of the Feynman kernels on
$\mathring{\mathbb{R}}^{n}$, where the weight factor $\chi(\sigma)$ is
a member of either the totally symmetric or totally antisymmetric
representations of $S_{n}$. This result has two distinct physical
meanings: the first is the existence of the Bose-Fermi alternative in
one dimension, which is expressed by the equations \eqref{eq:41a} and
\eqref{eq:41b}, and the second is the existence of the boson-fermion
duality in one dimension, which is expressed by the identity
\eqref{eq:42}. Then, by using \eqref{eq:42}, we have shown that---in
order for the Feynman kernel on $\mathcal{M}_{n}$ to satisfy the Robin
boundary conditions---the Feynman kernel on
$\mathring{\mathbb{R}}^{n}$ must satisfy the connection conditions for
the $\delta$-function ($\varepsilon$-function) potential if $\chi$ is
the totally (anti)symmetric representation. This proves the
boson-fermion duality between the systems described by the $n$-boson
Hamiltonian \eqref{eq:49a} and the $n$-fermion Hamiltonian
\eqref{eq:49b}, which are natural generalizations of the Lieb-Liniger
model \eqref{eq:1a} and the Cheon-Shigehara model \eqref{eq:1b},
respectively.

There are a number of directions for future work. One interesting
direction is to generalize the one-particle configuration space $X$ to
other geometries, such as the half-line
($X=\mathbb{R}/\mathbb{Z}_{2}$), a finite interval
($X=\mathbb{R}/D_{\infty}$), a circle ($X=\mathbb{R}/\mathbb{Z}$), and
graphs\footnote{For particle statistics on graphs, see
  \cite{Harrison:2010,Harrison:2013,Maciazek:2018,An:2020qyl,Maciazek:2020yee}.}. Because
the type and the number of particle statistics depends on $X$, one may
go beyond the boson-fermion duality by suitably choosing $X$. For
example, if there exist several particle statistics, one may have
triality, quadrality, pentality and so on rather than duality. Another
interesting direction is to generalize to the case in which
wavefunctions become multi-component. In this case, the weight factor
$\chi(\sigma)$ can be a higher-dimensional unitary representation of
the symmetric group $S_{n}$; that is, there can arise the
parastatistics rather than the Bose-Fermi alternative. In addition,
multi-component wavefunctions offer much more variety of two-body
contact interactions. For example, if wavefunctions have $N$
components, two-body contact interactions are generally described by
$U(N)$-family of boundary conditions. Note that such a generalization
can also be applied to many-body problems of non-identical
particles\footnote{The configuration space of non-identical particles
  in one dimension was discussed, e.g., in
  \cite{Loft:2014,Dehkharghani:2015,Harshman:2017}.}: if wavefunctions
have $N$ components, two-body contact interactions for distinguishable
particles are generally described by $U(2N)$-family of connection
conditions at the codimension-1 singularities. We hope to return to
all these things in the future.
      
Finally, let us comment on an implication for the possible coordinate
dependence on $a_{j}$. As noted in section \ref{section:2.2}, $a_{j}$
may depend on the coordinates parallel to the codimension-1
boundaries. This opens up a possibility to realize scale-invariant
two-body contact interactions which do not contain any dimensionful
parameters. A typical example for such coordinate dependence is given
by
\begin{align}
  a_{j}=g_{j}r,\quad j\in\{1,\cdots,n-1\},\label{eq:51}
\end{align}
where $g_{j}$ are dimensionless reals and $r$ is the hyperradius
defined in \eqref{eq:14}. Note that the hyperradius \eqref{eq:14} does
not contain the center-of-mass coordinate $\xi_{n}$ and is invariant
under the spatial translation $x_{j}\mapsto x_{j}+c$ for any
$c\in\mathbb{R}$. Hence in this case the Hamiltonians \eqref{eq:49a}
and \eqref{eq:49b} are also invariant under the spatial translation
such that the center-of-mass momenta are conserved. Note also that,
for $n\geq3$, the hyperradius $r$ is nonvanishing at the codimension-1
boundary $\partial\mathcal{M}^{\text{2-body}}_{n,j}$. In fact, $r$
vanishes only at the codimension-$(n-1)$ boundary
$\{x_{1}=x_{2}=\cdots=x_{n}\}$. Therefore, in the three- or more-body
problems of identical particles, we can construct scale-invariant
models without spoiling the translation invariance. Note, however,
that this continuous scale invariance could be broken down to a
discrete scale invariance in exactly the same way as the Efimov effect
\cite{Efimov:1970zz}. We will address this issue elsewhere.

\appendix
\section{Proof of the path-integral formula
  \texorpdfstring{\eqref{eq:33}}{(33)}}
\label{appendix:A}
Following the ideas presented in \cite{Ohya:2011qu,Ohya:2012qj}, in
this section we show that the formula \eqref{eq:33} satisfies the
properties \eqref{eq:27}--\eqref{eq:30} if $\chi$ is a one-dimensional
unitary representation of $S_{n}$ and if
$K_{\mathring{\mathbb{R}}^{n}}$ fulfills the conditions
\eqref{eq:34}--\eqref{eq:38}. Below we prove these four properties
separately.

\paragraph{Property 1.~(Composition law)}
Let us first prove the composition law \eqref{eq:27}. By substituting
\eqref{eq:33} into the left-hand side of \eqref{eq:27} we have
\begin{align}
  &\int_{\mathcal{M}_{n}}\!\!\!d\bm{z}\,K_{\mathcal{M}_{n}}(\bm{x},\bm{z};t_{1})K_{\mathcal{M}_{n}}(\bm{z},\bm{y};t_{2})\nonumber\\
  &=\sum_{\sigma\in S_{n}}\sum_{\sigma^{\prime}\in S_{n}}\chi(\sigma)\chi(\sigma^{\prime})\int_{z_{1}>\cdots>z_{n}}\!\!\!d\bm{z}\,K_{\mathring{\mathbb{R}}^{n}}(\bm{x},\sigma\bm{z};t_{1})K_{\mathring{\mathbb{R}}^{n}}(\bm{z},\sigma^{\prime}\bm{y};t_{2})\nonumber\\
  &=\sum_{\sigma\in S_{n}}\sum_{\sigma^{\prime}\in S_{n}}\chi(\sigma\sigma^{\prime})\int_{z_{1}>\cdots>z_{n}}\!\!\!d\bm{z}\,K_{\mathring{\mathbb{R}}^{n}}(\bm{x},\sigma\bm{z};t_{1})K_{\mathring{\mathbb{R}}^{n}}(\sigma\bm{z},\sigma\sigma^{\prime}\bm{y};t_{2})\nonumber\\
  &=\sum_{\sigma\in S_{n}}\sum_{\sigma^{\prime}\in S_{n}}\chi(\sigma\sigma^{\prime})\int_{w_{\sigma^{-1}(1)}>\cdots>w_{\sigma^{-1}(n)}}\!\!\!d\bm{w}\,K_{\mathring{\mathbb{R}}^{n}}(\bm{x},\bm{w};t_{1})K_{\mathring{\mathbb{R}}^{n}}(\bm{w},\sigma\sigma^{\prime}\bm{y};t_{2})\nonumber\\
  &=\sum_{\sigma^{\prime\prime}\in S_{n}}\chi(\sigma^{\prime\prime})\sum_{\sigma\in S_{n}}\int_{w_{\sigma^{-1}(1)}>\cdots>w_{\sigma^{-1}(n)}}\!\!\!d\bm{w}\,K_{\mathring{\mathbb{R}}^{n}}(\bm{x},\bm{w};t_{1})K_{\mathring{\mathbb{R}}^{n}}(\bm{w},\sigma^{\prime\prime}\bm{y};t_{2})\nonumber\\
  &=\sum_{\sigma^{\prime\prime}\in S_{n}}\chi(\sigma^{\prime\prime})\int_{\mathring{\mathbb{R}}^{n}}\!\!d\bm{w}\,K_{\mathring{\mathbb{R}}^{n}}(\bm{x},\bm{w};t_{1})K_{\mathring{\mathbb{R}}^{n}}(\bm{w},\sigma^{\prime\prime}\bm{y};t_{2})\nonumber\\
  &=\sum_{\sigma^{\prime\prime}\in S_{n}}\chi(\sigma^{\prime\prime})K_{\mathring{\mathbb{R}}^{n}}(\bm{x},\sigma^{\prime\prime}\bm{y};t_{1}+t_{2})\nonumber\\
  &=K_{\mathcal{M}_{n}}(\bm{x},\bm{y};t_{1}+t_{2}),\label{eq:A.1}
\end{align}
where in the second equality we have used the assumptions that $\chi$
is a representation that satisfies
$\chi(\sigma)\chi(\sigma^{\prime})=\chi(\sigma\sigma^{\prime})$ and
$K_{\mathring{\mathbb{R}}^{n}}$ satisfies the permutation invariance
\eqref{eq:38}. The third equality follows from the change of the
integration variable from $\bm{z}$ to $\bm{w}=\sigma\bm{z}$, the
fourth equality the change of the summation variables from $\sigma$
and $\sigma^{\prime}$ to $\sigma$ and
$\sigma^{\prime\prime}:=\sigma\sigma^{\prime}$, the fifth equality the
fact that the region $w_{\sigma^{-1}(1)}>\cdots>w_{\sigma^{-1}(n)}$
covers the whole $\mathring{\mathbb{R}}^{n}$ as $\sigma$ runs through
all possible permutations, and the sixth equality the assumption
\eqref{eq:34}. Hence we have shown that \eqref{eq:33} satisfies the
composition law \eqref{eq:27} if $\chi$ is a representation of $S_{n}$
and if $K_{\mathring{\mathbb{R}}^{n}}$ satisfies the permutation
invariance \eqref{eq:38} and the composition law \eqref{eq:34}.

\paragraph{Property 2.~(Initial condition)}
Let us next prove the initial condition \eqref{eq:28}. By substituting
\eqref{eq:33} into the left-hand side of \eqref{eq:28} we have
\begin{align}
  K_{\mathcal{M}_{n}}(\bm{x},\bm{y};0)
  &=\sum_{\sigma\in S_{n}}\chi(\sigma)K_{\mathring{\mathbb{R}}^{n}}(\bm{x},\sigma\bm{y};0)\nonumber\\
  &=\sum_{\sigma\in S_{n}}\chi(\sigma)\delta(\bm{x}-\sigma\bm{y})\nonumber\\
  &=\chi(e)\delta(\bm{x}-e\bm{y})\nonumber\\
  &=\delta(\bm{x}-\bm{y}),\label{eq:A.2}
\end{align}
where the second equality follows from the assumption
\eqref{eq:35}. In the third equality we have used the fact that, if
$\sigma$ is not the identity element $e$, $\bm{x}-\sigma\bm{y}$ cannot
be zero for any $\bm{x},\bm{y}\in \mathcal{M}_{n}$, which leads to
$\delta(\bm{x}-\sigma\bm{y})=0$ for $\sigma\neq e$. The last equality
follows from $\chi(e)=1$ and $e\bm{y}=\bm{y}$. Hence we have shown
that \eqref{eq:33} satisfies the initial condition \eqref{eq:28} if
$\chi$ is a representation of $S_{n}$ and if
$K_{\mathring{\mathbb{R}}^{n}}$ satisfies the initial condition
\eqref{eq:35}.

\paragraph{Property 3.~(Unitarity)}
Let us next prove the unitarity \eqref{eq:29}. By substituting
\eqref{eq:33} into the left-hand side of \eqref{eq:29} we have
\begin{align}
  \overline{K_{\mathcal{M}_{n}}(\bm{x},\bm{y};t)}
  &=\sum_{\sigma\in S_{n}}\overline{\chi(\sigma)}\,\,\overline{K_{\mathring{\mathbb{R}}^{n}}(\bm{x},\sigma\bm{y};t)}\nonumber\\
  &=\sum_{\sigma\in S_{n}}\chi(\sigma)^{-1}K_{\mathring{\mathbb{R}}^{n}}(\sigma\bm{y},\bm{x};-t)\nonumber\\
  &=\sum_{\sigma\in S_{n}}\chi(\sigma^{-1})K_{\mathring{\mathbb{R}}^{n}}(\bm{y},\sigma^{-1}\bm{x};-t)\nonumber\\
  &=K_{\mathcal{M}_{n}}(\bm{y},\bm{x};-t),\label{eq:A.3}
\end{align}
where the second equality follows from $\chi(\sigma)\in U(1)$ and the
assumption \eqref{eq:36}, the third equality
$\chi(\sigma)^{-1}=\chi(\sigma^{-1})$ and the assumption
\eqref{eq:38}. Hence we have shown that \eqref{eq:33} satisfies the
unitarity \eqref{eq:29} if $\chi$ is a one-dimensional unitary
representation of $S_{n}$ and if $K_{\mathring{\mathbb{R}}^{n}}$
satisfies the unitarity \eqref{eq:36} as well as the permutation
invariance \eqref{eq:38}.

\paragraph{Property 4.~(Schr\"{o}dinger equation)}
Let us finally prove that \eqref{eq:33} satisfies the Schr\"{o}dinger
equation \eqref{eq:30}. By substituting \eqref{eq:33} into the
left-hand side of \eqref{eq:30} we have
\begin{align}
  \left(i\hbar\frac{\partial}{\partial t}+\frac{\hbar^{2}}{2m}\bm{\nabla}_{\bm{x}}^{2}\right)K_{\mathcal{M}_{n}}(\bm{x},\bm{y};t)
  &=\sum_{\sigma\in S_{n}}\chi(\sigma)\left(i\hbar\frac{\partial}{\partial t}+\frac{\hbar^{2}}{2m}\bm{\nabla}_{\bm{x}}^{2}\right)K_{\mathring{\mathbb{R}}^{n}}(\bm{x},\sigma\bm{y};t)\nonumber\\
  &=0,\label{eq:A.4}
\end{align}
where we have used the assumption \eqref{eq:37}. Hence we have shown
that \eqref{eq:33} satisfies the Schr\"{o}dinger equation
\eqref{eq:30} if $K_{\mathring{\mathbb{R}}^{n}}$ satisfies the
Schr\"{o}dinger equation \eqref{eq:37}.

\bigskip

Now, as discussed in section \ref{section:3}, eq.~\eqref{eq:33} also
satisfies the Robin boundary conditions \eqref{eq:31} if
$K_{\mathring{\mathbb{R}}^{n}}$ satisfies the connection conditions
\eqref{eq:46}. Thus the formula \eqref{eq:33} fulfills all the
required properties \eqref{eq:27}--\eqref{eq:31} if $\chi$ is a
one-dimensional unitary representation of $S_{n}$ and if
$K_{\mathring{\mathbb{R}}^{n}}$ satisfies the assumptions
\eqref{eq:34}--\eqref{eq:38} as well as the connection conditions
\eqref{eq:46}.

\section{Proof of the integral formula
  \texorpdfstring{\eqref{eq:40}}{(40)}}
\label{appendix:B}
In this section we prove the integral formula \eqref{eq:40}. To this
end, let $f$ be an arbitrary test function on
$\mathring{\mathbb{R}}^{n}$. Then we have
\begin{align}
  \int_{\mathring{\mathbb{R}}^{n}}\!\!d\bm{y}\,f(\bm{y})
  &=\sum_{\sigma\in S_{n}}\int_{y_{\sigma(1)}>\cdots>y_{\sigma(n)}}\!\!\!d\bm{y}\,f(\bm{y})\nonumber\\
  &=\sum_{\sigma\in S_{n}}\int_{z_{1}>\cdots>z_{n}}\!\!\!d\bm{z}\,f(\sigma^{-1}\bm{z})\nonumber\\
  &=\int_{z_{1}>\cdots>z_{n}}\!\!\!d\bm{z}\left(\sum_{\sigma\in S_{n}}f(\sigma^{-1}\bm{z})\right),\label{eq:B.1}
\end{align}
where in the second line we have changed the integration variable as
$\bm{y}=\sigma^{-1}\bm{z}$. By changing the notations as
$\sigma^{-1}\to\sigma$ and $\bm{z}\to\bm{y}$ in the last line, we
arrive at the formula \eqref{eq:40}.

\printbibliography
\end{document}